\documentclass[apj]{emulateapj}
\usepackage{apjfonts}
\slugcomment{Accepted in Astrophys. J.}

\newcommand{\nuc}[2]{${}^{#2} \rm #1$}
\newcommand{\text}[1]{\rm #1}

\def\gtaprx {\lower .14ex\hbox{\rlap{\raise .9ex\hbox{\hskip .3ex
	{\ifmmode{\scriptscriptstyle >}\else
		{$\scriptscriptstyle >$}\fi}}}
	\kern -.4ex{\ifmmode{\scriptscriptstyle \sim}\else
		{$\scriptscriptstyle\sim$}\fi}}}
\def\ltaprx {\lower .14ex\hbox{\rlap{\raise .9ex\hbox{\hskip .3ex
	{\ifmmode{\scriptscriptstyle <}\else
		{$\scriptscriptstyle <$}\fi}}}
	\kern -.4ex{\ifmmode{\scriptscriptstyle \sim}\else
		{$\scriptscriptstyle\sim$}\fi}}}

\newcommand{\pp}{{\it p}}

\newcommand{\mev}{\, {\rm MeV} }

\newcommand{\g}{\, {\rm g} }
\newcommand{\cm}{\, {\rm cm} }
\newcommand{\s}{ \, {\rm s} }
\newcommand{\K}{ {\,\rm K} }

\newcommand{\cc}{\, {\rm cm^3} }
\newcommand{\erg}{\, {\rm erg} }

\newcommand{\km}{{\, \rm km}}
\newcommand{\ms}{{\, \rm ms}}

\newcommand{\gpccm}{{\, \rm g\,cm^{-3}}}

\newcommand{\nue}{\nu_{\rm e}} 
\newcommand{\nueb}{{\bar \nu}_{\rm e}} 
\newcommand{\nux}{\nu_x} 
\newcommand{\nuxb}{\bar \nu_x} 

\newcommand{\del}[2]%
{\frac{\mathrm{d}{#2}}{\mathrm{d}{#1}}}
\newcommand{\Del}[2]%
{\frac{D{#2}}{D{#1}}}\newcommand{\ddel}[2]%
{\frac{\mathrm{d}^2{#2}}{\mathrm{d}{#1}^2}}

\newcommand{\pdel}[2]%
{\frac{\partial{#2}}{\partial{#1}}}
\newcommand{\pddel}[2]%
{\frac{\partial^2{#2}}{\partial{#1}^2}}

\newcommand{\laplace}{\bigtriangleup}

\renewcommand{\vec}[1]{\mathbf{#1}}

\newcommand{\Ms}{M_{\odot}}
\newcommand{\DMs}{M_{\odot}\rm \,s^{-1}}

\shorttitle{Nucleosynthesis in neutrino-driven aspherical supernova explosions}
\shortauthors{Fujimoto et al.}

\begin{document}

\title{
Explosive nucleosynthesis in the neutrino-driven aspherical supernova
explosion of a non-rotating 15$M_{\odot}$ star with solar metallicity
}

\author{
Shin-ichiro Fujimoto\altaffilmark{1},
Kei Kotake\altaffilmark{2},
Masa-aki Hashimoto\altaffilmark{3},
Masaomi Ono\altaffilmark{3}, and
Naofumi Ohnishi\altaffilmark{4}
}

\altaffiltext{1}{
 Kumamoto National College of Technology, 
 2659-2 Suya, Goshi, Kumamoto 861-1102, Japan;
 fujimoto@ec.knct.ac.jp.
}

\altaffiltext{2}{
 Division of Theoretical Astronomy, 
 National Astronomical Observatory Japan, 
 2-21-1, Osawa, Mitaka, Tokyo, 181-8588, Japan.
}

\altaffiltext{3}{
 Department of Physics, School of Sciences, 
 Kyushu University, Fukuoka 810-8560, Japan.
}

\altaffiltext{4}{
 Department of Aerospace Engineering, Tohoku University, 
 6-6-01 Aramaki-Aza-Aoba, Aoba-ku, Sendai, 980-8579, Japan.
}

\begin{abstract}
We investigate explosive nucleosynthesis in a non-rotating 15$M_\odot$ star
with solar metallicity that explodes by a neutrino-heating supernova (SN) mechanism
aided by both standing accretion shock instability (SASI) and convection. 
To trigger explosions in our two-dimensional hydrodynamic simulations, 
we approximate the neutrino transport with a simple light-bulb scheme and systematically change
the neutrino fluxes emitted from the protoneutron star. 
By a post-processing calculation,  we evaluate abundances and masses of the SN ejecta for nuclei 
with the mass number $\le 70$ employing a large nuclear reaction network.
Aspherical abundance distributions, 
which are observed in nearby core-collapse SN remnants, 
are obtained for the non-rotating spherically-symmetric progenitor, due to the growth of low-mode SASI.
Abundance pattern of the supernova ejecta is similar to 
that of the solar system for models whose masses ranges $(0.4-0.5) \Ms$ 
of the ejecta from the inner region ($\le 10,000\km$) of the precollapse core.
For the models, the explosion energies and the \nuc{Ni}{56} masses are
$ \simeq 10^{51} \rm erg$ and $(0.05-0.06) \Ms$, respectively;
their estimated baryonic masses of the neutron star are comparable to 
the ones observed in neutron-star binaries.
These findings may have little uncertainty because
most of the ejecta is composed by matter that is heated via the shock wave 
and has relatively definite abundances.
The abundance ratios for Ne, Mg, Si and Fe observed in Cygnus loop 
are well reproduced with the SN ejecta from an inner region of the $15\Ms$ progenitor.
\end{abstract}


\section{Introduction}

The explosion mechanism of core-collapse supernovae (SNe) is still not clearly understood.
Multi-dimensional effects such as standing accretion shock instability (SASI)
and convection are recognized to be most important for unveiling the explosion mechanism, 
in particular for a progenitor heavier than about $11 M_{\odot}$ 
in its main sequence phase~\citep{kitaura06,buras06a,buras06b}.
Here SASI, becoming very popular in current supernova researches, 
is a uni- and bipolar sloshing of the stalled supernova shock with pulsational strong expansion and contraction 
(see, e.g.,\citet{blondin_03,scheck_04,ohnishi06,fogli07,blondin07a,iwakami08,iwakami09,nordhaus10} 
and references therein). 
Some of recent two-dimensional (2D) radiation-hydrodynamic simulations show 
that the delayed neutrino-driven mechanism aided by SASI and convection does work to produce
aspherical explosions \citep{marek09a,marek09b,suwa10}.

Observationally, global anisotropies and mixing as well as smaller-scale clumping of the SN ejecta, 
are common features of SN remnants like in SN1987A \citep{wang02}, 
Cas A~\citep{hughes00, willingale02}, G292.0+1.8~\citep{park07}, 
and Cygnus loop~\citep{kimura09, uchida09}. Asymmetries commonly
observed in the nebular emission-line profiles are considered as an evidence that
core-collapse SNe occur generally aspherically~\citep{maeda08, modjaz08, tanaka09b,taubenberger09}.
Evidences for asymmetry are also obtained from spectropolarimetric observations of 
Type Ibc SN at an early phase ($\sim$ days) (see, e.g.,\citet{tanaka08,tanaka09a} and references therein).

Thus far, nucleosynthesis studies of the SN ejecta
have almost successfully reproduced the solar composition and abundances
of radioactives observed in SN1987A \citep{hashimoto95, ww95, tnh96, rauscher02}.
However those spherical models have some problems such as 
overproduction of neutron-rich Ni isotopes and underproductions of \nuc{Ti}{44},
\nuc{Zn}{64} and light \pp-nuclei~\citep{rauscher02}. 

Aspherical effects on the explosive nucleosynthesis have been investigated by 
\citet{nagataki97, nagataki00}.
Based on 2D hydrodynamic simulations in which the
explosion was triggered by some form of manual energy deposition into a
stellar progenitor model outside the so-called mass cut, they evaluated
the composition of the ejecta with a large nuclear reaction network.
They pointed out that \nuc{Ti}{44} can be produced more abundantly in
the case of jet-like explosions compared to that of spherical explosions. 
\citet{young06} examined the composition of the ejecta
in three-dimensional (3D) SPH simulations to discuss a candidate of the progenitor of Cas A. 
They showed that the abundances of \nuc{Ni}{56} and \nuc{Ti}{44} 
depend on the magnitude and asymmetry of the explosion energy as well as on the amount of the fallback.
The effects of the fallback on the abundances have been systematically studied 
in one-dimensional explosion models~\citep{young07}. 
More recently, 3D effects have been more elaborately studied \citep{hunger03,hunger05}, 
as well as the impacts of different explosions by employing a number of progenitors~\citep{jog09,jog10}
or by assuming a jet-like explosion~\citep{couch09,tominaga09}, 
which is one of the possible candidates of hypernovae (e.g., \citet{maeda03,nagataki06}). 

In addition to the above-mentioned work, nucleosynthesis in a more realistic
simulation that models the multidimensional neutrino-driven 
SN explosion has been also extensively studied~\citep{kifonidis03, kifonidis06, gawryszczak10}.
Although a small network has ever been included in the computations, 
these 2D simulations employing a light-bulb scheme \citep{kifonidis03}
or a more accurate gray transport scheme \citep{scheck06,kifonidis06}
have made it possible to elucidate the nucleosynthesis inside from the iron core after the shock-revival
up to explosion in a more consistent manner.
\citet{kifonidis06} demonstrated that the SASI-aided low-mode explosions can most naturally explain the masses and
distribution of the synthesized elements observed in SN1987A.
Their recent 3D results by \citet{hammer10} show that 
the 3D effects that affect the velocity of the ejecta as well as the
growth of the Rayleigh-Taylor instability are really important to correctly determine the properties of the ejecta.

In the present work, we study explosive nucleosynthesis in a
non-rotating 15 $M_\odot$ star with solar metallicity by performing 2D hydrodynamic simulations
that models a SASI-aided delayed explosion via a light-bulb scheme. To extract a detailed information of the
synthesized elements, we follow the abundance evolution by employing a large
nuclear reaction network. It should be emphasized that the mass cut as
well as the aspherical distribution of the explosion energy 
are evaluated from the hydrodynamic simulations, as in our previous work on the nucleosynthesis 
in magnetohydrodynamically-driven SN explosions \citep{nishimura06}
as well as in collapsars \citep{fujimoto07,fujimoto08,ono09}.

In \S 2, this paper opens up with a brief description of a numerical code for
the hydrodynamic calculation, initial conditions of the progenitor star, and properties of the aspherical explosion. 
In \S 3, we present a large nuclear reaction network, 
physical properties of SN ejecta, 
and abundances and masses of the ejecta, and heavy-nuclei distribution of the SN ejecta.
We discuss the uncertainty in the estimate of the abundances and masses and 
compare the evaluated abundances with those  observed in Cygnus loop in \S 4.
Finally we will summarize our results in \S 5.

\section{Hydrodynamic simulations of an aspherical neutrino-driven supernova explosion}
\label{sec:hd_simulation}

\subsection{Hydrodynamic code and initial conditions}
\label{sec:hd_code}

To calculate the structure and evolution of the collapsing star, 
we solve the Newtonian hydrodynamic equations, 
\begin{equation}
\Del{t}{\rho}+\rho\nabla\cdot\vec{v}=0,
\end{equation}
\begin{equation}
\rho \Del{t}{\vec{v}}=-\nabla P -\rho \nabla (\Phi +\Phi_{c})
\end{equation}
\begin{equation}
 \rho \frac{d}{dt}\displaystyle{\Bigl( \frac{e}{\rho} \Bigr)}
  = - P \nabla \cdot \vec{v} + Q_{\rm E}, 
  \label{eq:energy}
\end{equation}
\begin{equation}
 \Del{t}{Y_e} = Q_{\rm N},
  \label{eq:ye_flow}
\end{equation}
where $\rho,P,\vec{v},e$, and $Y_e$, are the mass density, the pressure
the fluid velocity, the internal energy density, and the electron fraction, respectively.
We denote the Lagrange derivative as $D/Dt$. 
The gravitational potential of fluid and the central object with a mass of $M_{\rm in}$, 
$\Phi$ and $\Phi_{c}$, are evaluated with
\begin{equation}
\laplace{\Phi} = 4\pi G \rho,
\end{equation}
and
\begin{equation}
 \Phi_c = - \frac{G M_{\rm in}}{r},
\end{equation}
where $G$ is the gravitational constant.
We note that $M_{\rm in}$ continuously increases due to mass accretion through the inner boundary.

$Q_{\rm E}$ and $Q_{\rm N}$ are
the source terms that describe the rate of change per unit volume 
in equations (\ref{eq:energy}) and (\ref{eq:ye_flow}), respectively,
and will be summarized in Appendix A and B.
In the present study, we take into account absorption of electron and anti-electron neutrinos 
as well as neutrino emission through electron and positron captures, electron-positron pair annihilation, 
nucleon-nucleon bremsstrahlung, and plasmon-decays.
We assume that the fluid is axisymmetric and that
neutrinos are isotropically emitted from the neutrino spheres
with given luminosities and with the Fermi-Dirac distribution of given temperatures~\citep{ohnishi06}.
Rates for absorption of neutrinos and neutrino emission through electron and positron captures
are taken from \citet[appendix D]{scheck06}.
Geometrical factor $f_\nu$ is set to be 
\begin{equation}
  f_\nu \, = \, {\textstyle \frac{1}{2}} \Big[ 1+\sqrt{1-(R_\nu/r)^2} \;  \Big],
\end{equation}
as in \citet{scheck06}.
Here $R_\nu$ is the radius of neutrino sphere and is simply estimated with the relation, 
$L_{\nu} = \frac{7}{16}\sigma T_{\nu}^{4} \cdot 4\pi R_{\nu}^{2}$
for a given set of the luminosity $L_{\nu}$ and temperature $T_{\nu}$~\citep{ohnishi06}, 
where $\sigma$ is the Stefan-Boltzmann constant,
We adopt rates for the emission of neutrinos ($\nue,\nueb,\nux,\nuxb$) through pair annihilation, 
bremsstrahlung, and plasmon-decays as in \citet[Appendix B]{ruffert96}.
Moreover, we include the heating term in $Q_{\rm E}$ due to the absorption of neutrinos on \nuc{He}{4} and 
the inelastic scatterings on \nuc{He}{4} via neutral currents~\citep{haxton88,ohnishi07}.

The numerical code for the hydrodynamic calculations
employed in this paper is based on the ZEUS-2D code~\citep{sn92, ohnishi06}.
We use a realistic equation of state (EOS) based on the relativistic mean field theory~\citep{shen98}.
For lower density regime ($\rho < 10^5 \g/\cc$), where no data is available in the EOS table with the Shen EOS, 
we use another EOS, which includes contributions from an ideal gas of nuclei, radiation, and electrons and positrons
with arbitrary degrees of degeneracy~\citep{bdn96}.
We carefully connect two EOS at $\rho = 10^5 \g/\cc$ for physical quantities to vary continuous in density 
at a given temperature~\citep{fujimoto06}.

First we perform a spherical symmetric hydrodynamic simulation of 
the core collapse of a 15$M_\odot$ non-rotating star with the solar metallicity~\citep{ww95} 
using a hydrodynamic code~\citep{kotake04} for about $10\, \rm ms$ after core-bounce,
when the bounce shock turns into a standing accretion shock and the proto neutron star
 (PNS) grows to $\sim$ 1.2$M_\odot$.
Then, we map distributions of densities, temperatures, radial velocities 
and electron fractions of the spherical symmetric simulation to initial 
distribution for two-dimensional (2D) hydrodynamic simulations. 
After the remap, the central region inside 50 km in radius is excised to follow a 
long-term postbounce evolution (e.g., \citet{scheck06, kifonidis06}).
We impose velocity perturbations to the unperturbed radial velocity 
in a dipolar manner, and follow the postbounce evolution.
The spherical coordinates are used 
in our simulations and the computational domain is extended over
$50 \km \le r \le 50,000 \km$ and $0 \le \theta \le \pi$, or from the Fe core to inner O-rich layers, 
which are covered with 500($r$) $\times$ 128($\theta$) meshes.
The mass is $3.17\Ms$ in the computational domain.
We note that convective motion occurs at the onset of the 2D simulation 
with the above meshes, while the motion does not appear 
in the case of coarser mesh points of 500($r$) $\times$ 60($\theta$).
 Evolution of the explosion energy and mass ejection rate are very similar to those
for high resolution simulation with 300($r$) $\times$ 196($\theta$) meshes 
($50 \km \le r \le 3,000 \km$ and $0 \le \theta \le \pi$) for about 350 ms after the core bounce.
Therefore resolution of the simulations with 500($r$) $\times$ 128($\theta$) meshes seems to be 
appropriate for the present study.
However the resolution may be too low to follow later time evolution of
the explosion towards homologous expansion~\citep{gawryszczak10}.
For the high resolution simulation,
the minimum grid size in the radial- and lateral($\theta$)-directions, 
$\delta r$ and $\delta \theta$, is $1\km$ and $\pi/196$, respectively, 
while $\delta r = 1\km$ and $\delta \theta = \pi/128$ 
for our fiducial set (i.e., 500($r$) $\times$ 128($\theta$) mesh points).

\subsection{Aspherical SN explosion}
\label{sec:explosion}

We have performed the simulations for models with the electron-neutrino luminosities, 
$L_{\nu_e} = $3.7, 3.9, 4.0, 4.2, 4.5, 4.7, and 5.0 $\times 10^{52} \rm \, erg \, s^{-1}$
for $1-2\s$ after the core bounce, when a shock front has reached to a layer 
with $r = 10,000\km$ in almost all directions.
We take the input neutrino luminosities as above because 
the revival of the stalled bounce shock occurs only for models with $L_{\nu_e} \ge $3.9 $\times 10^{52} \rm \, erg~s^{-1}$, and also because for models with $L_{\nu_e} > 5.0 \times 10^{52} \rm \, 
erg~s^{-1}$, the star explodes too early for the SASI to grow, as will be discussed later.
We set $L_{\bar{\nu_e}} = L_{\nu_e}$ and $L_{\nu_x} = 0.5 L_{\nu_e}$,
where $L_{\bar{\nu_e}}$ and $L_{\nu_x}$ are the luminosities of anti-electron neutrino and other-types 
($\mu$, $\tau$, anti-$\mu$, and anti-$\tau$), respectively.
We consider models with neutrino temperatures, $T_{\nu_e}$, $T_{\bar{\nu_e}}$, and $T_{\nu_x}$ 
as $4 \, \rm MeV$, $5 \, \rm MeV$, and $10 \, \rm MeV$, respectively~\citep{ohnishi06}.
The adopted neutrino luminosities are comparable to those with a more accurate transport scheme
but the temperatures are slightly higher~\citep{marek09a,marek09b}.
We will present hydrodynamic and nucleosynthetic results for cases with lower neutrino temperatures,
in \S \ref{sec:neutrino-temperature}.

\begin{figure}[ht]
 \epsscale{0.45} 
 \plotone{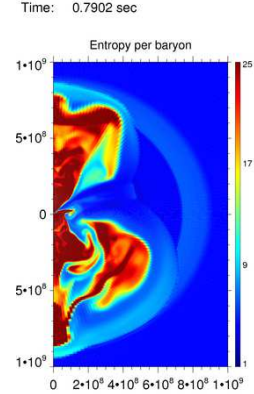}
 \caption{Entropy contour for case with $L_{\nu_e} = 4.5 \times 10^{52} \rm \, erg \, s^{-1}$ 
at 0.79s after the core bounce. 
The growth of low-mode SASI is prominent.
}
\label{fig:entropy-hE}
\epsscale{1.0}
\end{figure}

We confirm that the explosion are highly aspherical and $l=1$ and $l=2$ modes are dominant
as shown in \citet{kifonidis06, ohnishi06, scheck06}, 
although the shape of the explosion strongly depends on numerical detail, 
such as mesh resolution and boundary conditions~\citep{kifonidis06, scheck06}.
Entropy contour of the hydrodynamic simulation is shown in Figure \ref{fig:entropy-hE}
for case with $L_{\nu_e} = 4.5 \times 10^{52} \rm \, erg \, s^{-1}$.
Most of SN ejecta have entropy less than 20$k_{\rm B}$, where $k_{\rm B}$ is the Boltzmann constant.
Entropy attains to $70 k_{\rm B}$ for small amounts of the ejecta.

\begin{figure}[ht]
 \plotone{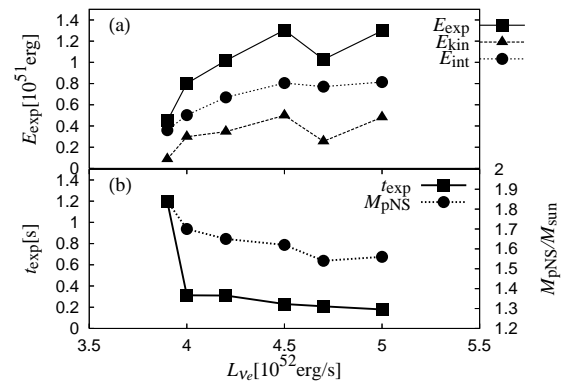}
 \caption{
 (a) Explosion energies vs. neutrino luminosities. 
 Filled squares, circles, and triangles indicate 
 explosion energies, kinetic-part and thermal-part of the explosion energies, respectively.
 (b) $M_{\rm pNS}$ and $t_{\rm exp}$ vs. neutrino luminosities. 
 Filled squares and circles indicate $t_{\rm exp}$ and $M_{\rm pNS}$, respectively.
}
 \label{fig:Ep-Lnu}
 \epsscale{1.0}
\end{figure}

We find that for models with $L_{\nu_e} \ge 3.9 \times 10^{52} \rm \, erg \, s^{-1}$,
the star explodes aspherically via the neutrino heating aided by SASI.
Figure \ref{fig:Ep-Lnu}(a) shows explosion energies as a function of $L_{\nu_e}$, for all the exploded models.
The energies are estimated at an epoch of 500 ms after the explosion and slightly increase after the epoch.
Kinetic and thermal energies of the explosion are also shown in Figure \ref{fig:Ep-Lnu}(a).
The thermal energies dominate over the kinetic ones.

Higher $L_{\nue}$ makes the onset of explosion earlier and also the mass 
of the PNS ($M_{\rm pNS}$) smaller. Note that we estimate
$M_{\rm pNS}$ at $t = t_{\rm exp} +500\ms$, because $M_{\rm pNS}$ only slightly 
increases later than 500 ms after $t_{\rm exp}$. Here $t_{\rm exp}$ indicates 
the time scale when the explosion sets in, which can be typically estimated
when the mass ejection rate at 100km grows up $0.1\DMs$ in our 2D simulations.
Figure \ref{fig:Ep-Lnu}(b) shows the explosion time $t_{\rm exp}$ and $M_{\rm pNS}$ as a function of $L_{\nue}$.
Except for the lowest luminosity model 
($L_{\nu_e} \le 3.9 \times 10^{52} \rm \, erg \, s^{-1}$), $M_{\rm pNS}$ is in the range of $1.54-1.70\Ms$.
These values are much larger than the so-called mass-cut in the spherical model 
of $15\Ms$ progenitors by \citet{hashimoto95} and \citet{rauscher02}, which is $1.30\Ms$ and $1.32\Ms$, respectively.
The mass of the PNS however becomes larger ($1.68\Ms$) due to the fallback of ejecta~\citep{rauscher02}.

\section{Nucleosynthesis in supernova ejecta}
\label{sec:nucleosynthesis}

\subsection{Nuclear reaction network and initial composition}
\label{sec:network}

In order to calculate chemical composition of the SN ejecta, 
we need Lagrangian evolution of physical quantities, such as density, temperature, and, velocity of the material.
We adopt a tracer particle method~\citep{nagataki97, seitenzahl10}
to calculate the Lagrangian evolution of the physical quantities 
from the Eulerian evolution obtained from our simulations.
The Lagrangian evolution is followed during the 2D aspherical simulation as well as the spherical collapsing phase.
To get information on mass elements, 6,000 tracer particles are placed in the 
regions from 300 extending to 10,000 km (the O-rich layer).
We have confirmed that the estimated energies and masses of the ejecta with the 6,000 particles 
are equal to the ones with 3,000 particles within $\sim$ 1\% accuracy and 
the obtained abundance profiles are also very similar.

Initial abundances of the particles are set to be those of 
the star just before the core collapse~\citep{rauscher02}, 
in which 1400 nuclei are taken into account.
We note that a presupernova model in \citet{rauscher02} has 
smaller helium, carbon-oxygen, and oxygen-neon core masses, compared with those in \citet{ww95},
due to coupled effects through the inclusion of mass loss and 
the revisions of opacity and nuclear inputs~\citep{rauscher02}.
The mass of a particle in a layer is weighted to the mass in the layer.
We note that the minimum mass of the particles is $\sim 10^{-4} \Ms$.
We find that more than one fifth particles are ejected due to the aspherical explosion.

Next we calculate abundances and masses of the supernova ejecta.
Ejecta that is located on the inner region of the star ($r_{\rm ej,cc} \le 10,000\km$)
before the core collapse, has high maximum temperatures 
enough for elements heavier than C to burn explosively.
Here $r_{\rm ej,cc}$ is the radius of the ejecta at the core collapse.
We therefore follow abundance evolution of the ejecta from the inner region
using a nuclear reaction network,
which includes 463 nuclide from neutron, proton to Kr~\citep{fujimoto04}.
We will discuss effects of neutrino interactions on heavy nuclei and uncertainty in nuclear reaction rates 
on nucleosynthetic results, in \S \ref{sec:neutrino-effects} and \ref{sec:reaction-rates}, respectively.
While the abundances of ejecta from the outer region ($r_{\rm ej,cc} > 10,000\km$)
are set to be those before the core collapse~\citep{rauscher02}.
We note that the masses of the outer region, or $r_{\rm ej,cc} > 10,000\km$, is 10.4$\Ms$.
Moreover, when temperatures of the ejecta are greater than $9 \times 10^9 \K$, 
we set chemical composition of the ejecta to be that in nuclear statistical equilibrium (NSE), 
whose abundances are expressed with simple analytical expressions, 
specified by the density, temperature and electron fraction.

Electron fractions of the ejecta are re-evaluated during SN explosion coupled with the nuclear reaction network. 
The change in $Y_e$ is taken into account through electron and positron captures on heavy nuclei, 
in addition to electron and positron captures on neutrons and protons
as well as absorption of $\nue$ and $\nueb$ on neutrons and protons. 
The captures and absorptions on neutrons and protons are also taken into account in the hydrodynamic simulations.
The rates for the captures and the absorptions are adopted from \citet{ffn80,ffn82} and \citet{scheck06}, respectively.

It should be emphasized that post-processing electron fractions are slightly different (up to 10\%)
from those estimated with hydrodynamic simulations, in which the evolution of electron fractions is followed.
This is because abundances of neutrons and protons in the network calculations are slightly
different from those estimated with EOS in the hydrodynamic simulations. 
We note that 463 nuclei are taken into account in the network calculations, 
while only neutrons, protons, \nuc{He}{4}, and a representative heavier nuclide
are evaluated with EOS.

In the neutrino-heating dominated region, 
the abundances of neutrons and protons in the network calculations 
are larger than those evaluated with EOS.
Hence, if we perform hydrodynamic simulations, in which 
abundances of nucleons are reliably evaluated with the reaction network,
the neutrino heating rates in the simulations could increase compared to those in the current study, 
since the neutrino heating through the absorption of $\nue$ and $\nueb$ is dominant over 
the other heating reactions,
and the heating rates via the absorption are proportional to the abundances of the nucleons.

The explosion energies also might increase. We emphasize that abundances of SN ejecta chiefly depends
on the explosion energy and the mass of SN ejecta from an inner region, not on $L_{\nue}$, as shown in later.

\subsection{Physical properties of SN ejecta}
\label{sec:property}

\begin{figure}[ht]
 \plotone{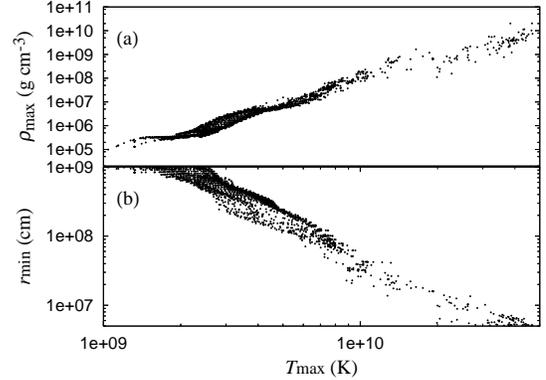}
 \caption{
 (a) Maximum densities vs. maximum temperatures of ejecta for 
  model with $L_{\nu_e} = 4.5 \times 10^{52} \rm \, erg \, s^{-1}$.
 (b) Minimum radial position, $r_{\rm min}$ vs. maximum temperatures of ejecta for 
 model with $L_{\nu_e} = 4.5 \times 10^{52} \rm \, erg \, s^{-1}$.
Ejecta that falls near the proto neutron star have high densities $>10^9 \gpccm$ 
and temperatures $> 10^{10} \K$.
 }
\label{fig:peak}
\epsscale{1.0}
\end{figure}
Maximum densities, $\rho_{\rm max}$, and maximum temperatures, $T_{\rm max}$, are good indicators 
for the composition of SN ejecta~\citep{tnh96}.
Figure \ref{fig:peak}(a) shows $\rho_{\rm max}$ as a function of $T_{\rm max}$
of the ejecta for $L_{\nu_e} = 4.5 \times 10^{52} \rm \, erg \, s^{-1}$.
Most of the ejecta with relatively low densities ($< 10^8 \gpccm$) 
have $\rho_{\rm max}$ and $T_{\rm max}$ similar to 
those of ejecta in the spherical model of core collapse SNe~\citep{thielemann98}.
For some particles that have very high densities simultaneously with high temperatures 
($\rho_{\rm max} \ge 10^{9} \gpccm$ and $T_{\rm max} \ge 10^{10} \K$), 
 electron captures operate to some extent, so that these particles become slightly 
neutron-rich, $Y_e\rm (10,000\km) < 0.48$.
Here $Y_e\rm (10,000\km)$ represents the electron fraction for tracer particles 
evaluated when the particles reach $r=10,000\km$. This may be a useful 
 quantity to measure $Y_e$ of the ejecta, since $Y_e$ closely freezes 
out at $r > 10,000\km$ (except for through $\beta$-decays at a later epoch).
All ejecta with $T_{\rm max} > 10^{10}\K$ falls down to the heating region $\le 200-300\km$
to be heated via neutrinos (Fig.\,\ref{fig:peak}(b)).
For these neutrino-heated ejecta, $\rho_{\rm max}$ range from $10^8 - 2 \times 10^{10}\gpccm$.

\begin{figure}[ht]
 \epsscale{0.8} 
 \plotone{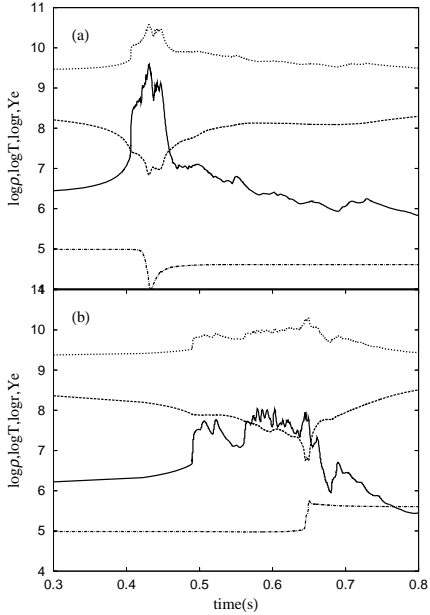}
 \caption{Time evolution of density, temperature, radial position, and electron fraction 
 of an ejecta with (a) $Y_e\rm (10,000\km) = 0.461$ and (b) $Y_e\rm (10,000\km) = 0.559$, 
 for model with $L_{\nu_e} = 4.5 \times 10^{52} \rm \, erg \, s^{-1}$.
 Solid, dotted, dashed, and dash-dotted lines represent
 the density, temperature, radial position, and electron fraction of the ejecta, respectively.
 The electron fraction is shown with a value multiplied by ten.
 The density, temperature, and the radial position of the ejecta are presented 
 in units of $\gpccm$, $\K$, and $\cm$, respectively.
 $Y_e$ of ejecta can largely change only in an inner region near the proto neutron star.
}
\label{fig:phys-evol-pe}
\epsscale{1.0}
\end{figure}

For ejecta with higher $\rho_{\rm max}$, electron captures on protons proceed 
more efficiently to make its $Y_e$ smaller.
Time evolution of physical quantities of such an ejecta is shown 
during the infall of the ejecta near the cooling region ($r < 100\km$) in Figure \ref{fig:phys-evol-pe}(a).
As the density and temperature rise to more than $10^9\gpccm$ and $10^{10}\K$, respectively, 
the electron fraction decreases due to the electron captures.
When the ejecta starts to be released via neutrino heating at $t = 0.42\s$, the electron fraction
increases through the absorption of $\nu_e$ by neutrons.
Finally $Y_e\rm (10,000\km)$ becomes 0.461 for the ejecta.

On the other hand, the electron captures on protons are not efficient for 
a proton-rich ejecta with $Y_e\rm (10,000\km) = 0.559$, 
as shown in Figure \ref{fig:phys-evol-pe}(b).
This is because the densities of the inner region ($r \le 200\km$) are relatively low ($\le 10^8 \gpccm$)
due to the mass ejection during an earlier phase ($\ge t_{\rm exp}$).
The electron fraction therefore remains constant and rises from 0.5 to 0.559 via the $\nu_e$ absorption 
in an inner region $r \le 200\km$.
The proton-richness in the ejecta is caused by the small energy difference between $\nue$ and $\nueb$.
For $T_{\nue}$ and $T_{\nueb}$ adopted in our simulations, 
the relation, $4(m_n -m_p) > \epsilon_{\nueb} -\epsilon_{\nue}$, holds, 
which leads to $Y_e > 0.5$~\citep{frohlich06a}, 
where $m_n$ and $m_p$ are masses of neutron and protons, and
$\epsilon_{\nueb}$ and $\epsilon_{\nue}$ are energies of anti-electron and electron neutrinos, respectively.
We note that $\epsilon_{\nueb} = 15.8\mev$ for $T_{\nueb} = 5\mev$ 
and $\epsilon_{\nue}= 12.6\mev$ for $T_{\nue} = 4\mev$.

\begin{figure}[ht]
 \plotone{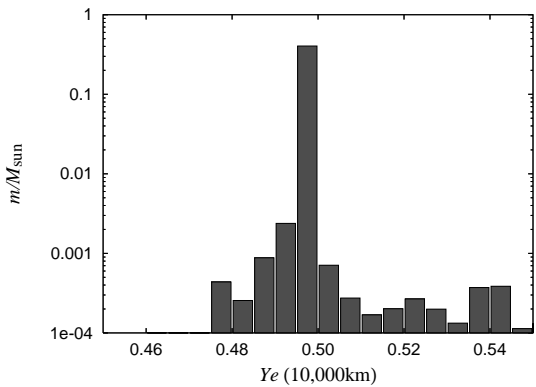}
 \caption{Masses as a function of $Y_e \rm (10,000km)$ of ejecta 
from the inner region $r_{\rm ej,cc} \le 10,000\km$
for model with $L_{\nu_e} = 4.5 \times 10^{52} \rm \, erg \, s^{-1}$.
}
\label{fig:Mye}
\epsscale{1.0}
\end{figure}

Figure \ref{fig:Mye} shows masses as a function of $Y_e \rm (10,000km)$ of ejecta 
from the inner region $r_{\rm ej,cc} \le 10,000\km$. 
We find that most of the ejecta (98.8\%) have electron fractions of 0.49-0.5.
Small fractions of the ejecta, 0.9\% and 0.3\% in mass, are slightly neutron-rich ($0.46<Y_e < 0.49$) 
and proton-rich ($0.5 < Y_e <0.56$), respectively.
Masses of the slightly neutron- and proton-rich ejecta are larger for models with larger $L_\nu$,
while the mass fractions of these ejecta are comparable for all the models.

\subsection{Primary and secondary ejecta}
\label{sec:pe-se}

\begin{figure}[ht]
 \plotone{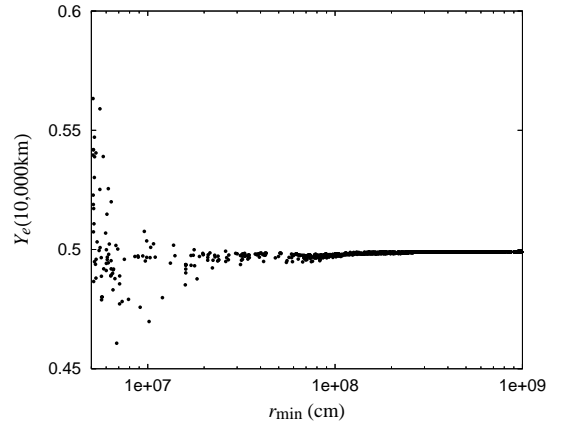}
 \caption{
 $Y_e\rm (10,000km)$ vs. minimum radial position, $r_{\rm min}$, of ejecta for 
  model with $L_{\nu_e} = 4.5 \times 10^{52} \rm \, erg \, s^{-1}$.
 $Y_e\rm (10,000km)$ are largely different from 0.5 only for ejecta that fall near the proto neutron star.
 }
\label{fig:ye-rmin}
\epsscale{1.0}
\end{figure}
Electron fraction of an ejecta with the minimum radial position $r_{\rm min} \le 200-300\km$ changes 
due to high neutrino flux and/or efficient e$^\pm$ capture (Fig.\,\ref{fig:phys-evol-pe}(a)).
Figure \ref{fig:ye-rmin} shows $Y_e \rm (10,000km)$ as a function of $r_{\rm min}$
of the ejecta for model with $L_{\nu_e} = 4.5 \times 10^{52} \rm \, erg \, s^{-1}$.
Hereafter, we refer to the ejecta with $r_{\rm min} \le 200\km$ as the {\itshape primary ejecta}, 
which have high maximum densities $\ge 10^8\gpccm$ and temperatures $\ge 10^{10}\K$ (Fig.\,\ref{fig:peak}).
The ejecta are heated through the neutrino heating.
On the other hand, the others are refereed as the {\itshape secondary ejecta}, 
heated chiefly via the shock wave driven by the primary ejecta.

It is true that $Y_e$ of the primary ejecta 
could change if we adopt a more accurate neutrino-transfer scheme 
instead of the simplified light-bulb transfer scheme. 
Abundances of the primary ejecta are therefore highly uncertain because of
the uncertainty on their $Y_e$.
On the other hand, for the secondary ejecta, $Y_e$ changes chiefly through the neutrino absorptions 
but the changes in $Y_e$ are found to be less than 1\% for almost all the secondary ejecta.
In addition, for our typical 2D models that produce energetic explosions 
($L_{\nu_e} \ge 4.5 \times 10^{52} \rm \, erg \, s^{-1}$), 
the masses of the primary ejecta occupy only about 2\% ($8.7 \times 10^{-3}\Ms$)
in the ejecta from the inner region (0.41$\Ms$ for $r_{\rm ej,cc} \le 10,000\km$).
Therefore, $Y_e$ of the secondary ejecta, 
whose mass is much larger than that of the primary ejecta, 
are unlikely to be largely changed even if we use a more accurate neutrino-transfer scheme.
We conclude that masses of abundant nuclei, such as \nuc{O}{16}, \nuc{Si}{28}, and \nuc{Ni}{56}, 
do not largely change in the SN ejecta.
We will discuss this point in \S \ref{sec:uncertainty}.

\begin{figure}[ht]
 \plotone{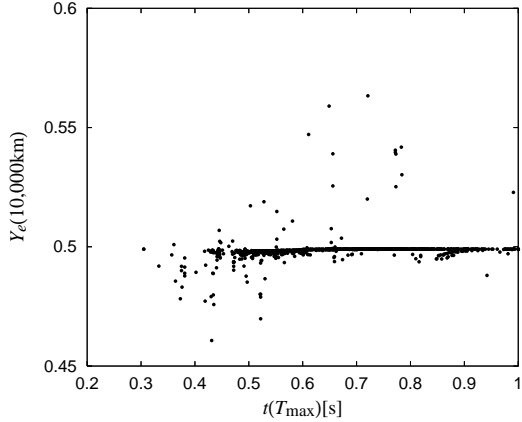}
 \caption{
 $Y_e$ vs. $t(T_{\rm max})$, 
 the time when the temperature of ejecta attains to $T_{\rm max}$ 
 for model with $L_{\nu_e} = 4.5 \times 10^{52} \rm \, erg \, s^{-1}$.
Ejecta at an early phase ($< 200-300\ms$ after the explosion) tend to be neutron-rich, 
while proton-rich in the later phase.
}
\label{fig:ye-t-Tmax}
\epsscale{1.0}
\end{figure}
The value of $Y_e$ of the primary ejecta depends on the epoch of the ejection.
The primary ejecta that eject in an early phase (before an epoch of $200-300\ms$ after the explosion 
($t_{\rm ej} = 230\ms$))
are mainly neutron-rich, while the primary ejecta are proton-rich in the later phase.
The proton-rich primary ejecta corresponds a kind of neutrino-driven winds, which are possibly 
not neutron-rich but proton-rich~\citep{fischer10, hudepohl10}.
Figure \ref{fig:ye-t-Tmax} shows $Y_e \rm (10,000km)$
as a function of $t(T_{\rm max})$ of ejecta for model 
with $L_{\nu_e} = 4.5 \times 10^{52} \rm \, erg \, s^{-1}$, 
where $t(T_{\rm max})$ is defined as the time when the temperature of ejecta attains to their maximum value, 
$T_{\rm max}$.
We note that the gas starts to be ejected just after $t = t(T_{\rm max})$, 
as shown in Figure \ref{fig:phys-evol-pe}.
We note that the dependence of $Y_e$ on the epoch of the ejection 
also appears in a spherical simulation of SN explosion of a star with an ONeMg core using an elaborated code
taking into account an accurate neutrino transfer scheme~\citep{kitaura06,wanajo09}.

\subsection{Masses and abundances of ejecta}
\label{sec:mass_abundance}

\begin{figure}[ht]
 \plotone{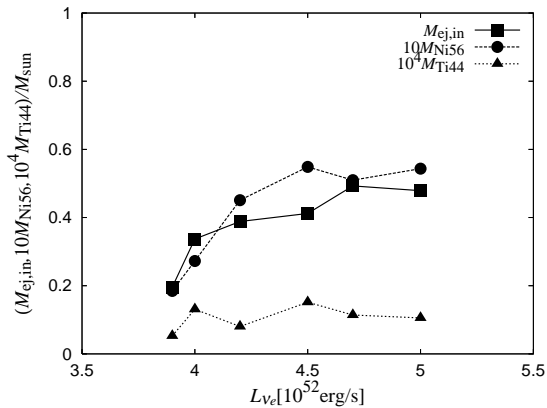}
 \caption{
 Mass of the ejecta from the inner region
 and masses of radioactives vs. $L_{\nu_e}$.
 Note in a spherical model by \citet{rauscher02} that the obtained 
masses of \nuc{Ni}{56} and \nuc{Ti}{44} are 
 0.11$M_{\odot}$ and $1.4 \times 10^{-5}\Ms$, respectively.
 }
 \label{fig:mass}
 \epsscale{1.0}
\end{figure}
Masses of nuclei, such as \nuc{Ni}{56}, \nuc{Ni}{57}, \nuc{Ni}{58} and \nuc{Ti}{44}, 
have been estimated in some SN remnants; 
For SN1987A, masses of \nuc{Ni}{56} and \nuc{Ti}{44} are deduced to be $\simeq 0.07\Ms$~\citep{shigeyama88, woosley88}
and $1-2\times 10^{-4}\Ms$~\citep[and references therein]{nagataki00}, respectively.
The estimated mass of \nuc{Ti}{44} is comparable to that in Cas A 
($1.6^{+0.6}_{-0.3} \times 10^{-4}\Ms$)~\citep{renaud06} 
and greater than that in the youngest Galactic supernova remnant G1.9+0.3 
($1-7 \times 10^{-5}\Ms$)~\citep{borkowski10}, which may originate from a Type Ia event.
Figure \ref{fig:mass} shows masses of SN ejecta ejected from the inner region ($r_{\rm ej,cc} \le 10,000\km$), 
$M_{\rm ej,in}$, and masses of \nuc{Ni}{56} and \nuc{Ti}{44}.
Masses of the ejecta from the inner region of $r_{\rm ej,cc} \le 10,000\km$ (solid line with filled squares) and
masses of \nuc{Ni}{56} (dashed line with filled circles) and \nuc{Ti}{44} (dotted line with filled triangles)
are shown with a value times a factor of 1, 10 and $10^4$, respectively.
We find that masses of the ejecta and \nuc{Ni}{56} 
roughly correlate with the neutrino luminosities.
The masses of \nuc{Ni}{56} and \nuc{Ti}{44} are less than and
comparable to those in the spherical model~\citep{rauscher02}, 
or 0.11$M_{\odot}$ and $1.4 \times 10^{-5}\Ms$, respectively.
Note that the explosion energy is $1.2 \times 10^{51}\erg$ in the model.
We find that \nuc{Ti}{44} relative to \nuc{Ni}{56} 
are much smaller than those in the solar system, in SN1987A, and in Cas A,
but comparable to that in the supernova remnant G1.9+0.3.
\nuc{Ni}{57} relative to \nuc{Ni}{56} is comparable to those in the solar system and in SN1987A, 
while \nuc{Ni}{58} relative to \nuc{Ni}{56} is overproduced compared with those in the solar system and in SN1987A, 
because they are abundantly produced in the slightly neutron-rich ejecta~\citep{hashimoto95,nagataki97}.

It should be emphasized that \nuc{Ti}{44} is underproduced in our simulations of the SN explosion, 
contrary to the overproduction in the previous 2D results \citep{nagataki97}, 
in which the explosion energy is aspherically and artificially added 
and the remnant mass is set to a value in order that the ejected mass of \nuc{Ni}{56} is reproduced 
the mass observed in SN1987A in \citet{nagataki97}.
The underproduction is possibly caused by a larger remnant mass in our models.
This is because the explosion energy and the ratio of the explosion energy on the polar axis 
to that on the equatorial plane are comparable to those evaluated in our simulation, 
and masses of \nuc{Ti}{44} as well as \nuc{Ni}{56} have shown to 
strongly depend on the value of the remnant mass in 2D calculations~\citep{young06}.

\begin{figure}[ht]
 \plotone{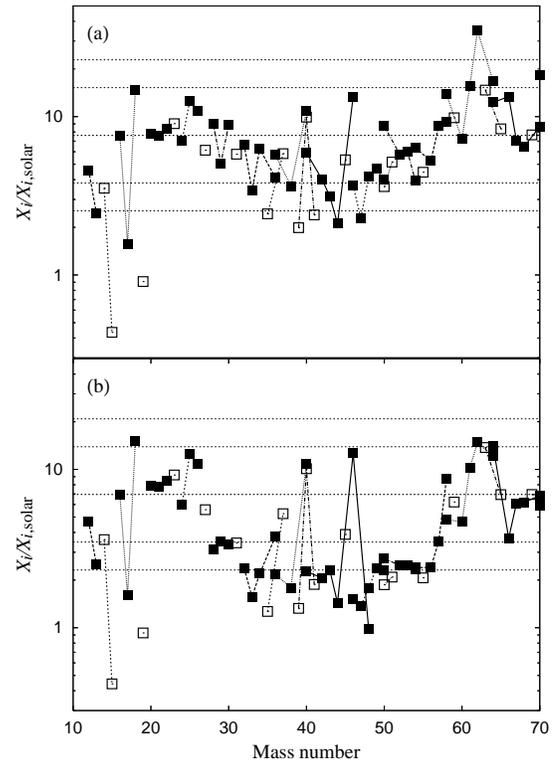}
 \caption{
 Overproduction factors, $X_i/X_{\odot,i}$ vs. mass number
 for (a) $L_{\nu_e} = 4.5 \times 10^{52} \rm \, erg \, s^{-1}$
 and for (b) $L_{\nu_e} = 3.9 \times 10^{52} \rm \, erg \, s^{-1}$
 Thick horizontal-dashed lines represents a factor equals to that of \nuc{O}{16},
 while two normal and two thin lines denote
 a factor equals to that of \nuc{O}{16} times 2, 1/2, 3, and 1/3, respectively.
 }
 \label{fig:opf}
 \epsscale{1.0}
\end{figure}
In order to compare estimated abundances with the solar system ones~\citep{anders89},
we have integrated masses of nuclei over all the ejecta to evaluate the abundances of the SN ejecta.
Figure \ref{fig:opf}(a) shows overproduction factors after decays as a function of the mass number, $A$
for $L_{\nu_e} = 4.5 \times 10^{52} \rm \, erg \, s^{-1}$,
in which the explosion energy is 1.3 $\times 10^{51} \rm \, erg \, s^{-1}$ (Fig.\,\ref{fig:Ep-Lnu}(a))
and $M_{\rm ej,in}$ is $0.41\Ms$ (Fig.\,\ref{fig:mass}).
We find that the abundance pattern of the SN ejecta is similar to 
the solar system one. 
We note that \nuc{O}{17}, which is underproduced in the ejecta, 
can be abundantly synthesized in Type Ia SNe,
and that \nuc{N}{15} and \nuc{F}{19} are comparably produced if neutrino effects are taken into account~\citep{ww95}, 
as shown later in Figure \ref{fig:comp-neutrino}. 
We point out that \nuc{Zn}{64}, which is underproduced in the spherical case~\citep{rauscher02}, 
is abundantly produced in slightly neutron-rich ejecta ($0.46 \le Y_e \le 0.49$).

In the ejecta, however, the neutron-rich Ni, \nuc{Ni}{62}, is overproduced.
The overproduction of the neutron-rich Ni isotopes has also appeared in the spherical models~\citep{hashimoto95,rauscher02}
as well as in the asymmetric model~\citep{nagataki97}.
The overproduction has been suppressed if the electron fraction of slightly neutron-rich material 
($Y_e = 0.495$ just before the core collapse) near the mass cut is artificially modified to 
0.499~\citep{hashimoto95,nagataki97}.
Hence, it has been remarked that the overabundance of the neutron-rich Ni isotopes
may inherit the uncertainty in the progenitor model,
in particular in the neutronization due to the combined effects of convective mixing,
electron captures, and positron decays during their Si burring stage~\citep{hashimoto95}.
In our aspherical models, 
the overabundances of Ni in the primary neutron-rich ejecta are not large, other than \nuc{Ni}{60}, 
as we will discuss later (\S \ref{sec:uncertainty}).
The overproduction of Ni mainly takes place in slightly neutron-rich secondary ejecta ($0.49 < Y_e < 0.5$).
We find that the electron fractions of the secondary ejecta are 0.4985 just before the core collapse 
and decrease by up to 1\% through the neutrino absorptions.
Therefore, the overabundance of \nuc{Ni}{62} possibly inherits the uncertainty
not only in the progenitor model but also in the change for $Y_e$ during the SN explosion.
Moreover, the overproduction of Ni isotopes are also shown to 
be reduced in the ejecta from the spherical SN explosion 
induced via artificially enhanced neutrino heating~\citep{frohlich06a}.

On the other hand, the integrated abundances are largely different from those in the solar system
for $L_{\nu_e} = 3.9 \times 10^{52} \rm \, erg \, s^{-1}$, in which
$E_{\rm exp} = 0.45 \times 10^{51} \rm \, erg \, s^{-1}$ and $M_{\rm ej,in} = 0.20\Ms$.
Nuclei with $A \ge 28$ are deficient in the model (Fig.\,\ref{fig:opf}(b)), 
compared with O, which is mainly ejected from outer layers.
The overproduction factors of nuclei lighter than \nuc{Al}{27} are very similar for both models,
because these nuclei are mainly synthesized during the hydrostatic burning.
While nuclei heavier than Si are mainly produced during the SN explosion through the explosive burning
in ejecta from the inner region ($r_{\rm ej,cc} \le 10,000\km$).

The abundance patterns for models with $M_{\rm ej,in} \sim (0.4-0.5)\Ms$, or 
those with $L_{\nu_e} = $ (4.5, 4.7, and 5.0) $\times 10^{52} \rm \, erg \, s^{-1}$, 
are similar to that in the solar system.
We should notice that $M_{\rm ej,in}$, which corresponds to masses of a neutron star,
estimated as $(1.54-1.62) \Ms$ for these models (Fig.\,\ref{fig:Ep-Lnu}(b)), 
are comparable to the baryonic mass (around $1.5\Ms$) of a neutron star
for observed neutron-star binaries, 
in which the progenitor of the neutron star seems to be a star with a Fe core~\citep{schwab10}.
Moreover, $E_{\rm exp}/M_{\rm ej} = (0.92 - 1.2) \times 10^{50} \erg/\Ms$ evaluated for the models
are comparable to the value $E_{\rm exp}/M_{\rm ej} = 0.76 \times 10^{50} \erg/\Ms$ 
estimated in SN 1987A~\citep{shigeyama90}.
The time from the core collapse to the explosion, $t_{\rm exp}$, 
is $\sim 0.2\s$ (Fig.\,\ref{fig:Ep-Lnu}(b)), which is enough to grow a low-mode SASI.
The growth is appropriate for explaining the distributions and velocities of nuclei
observed in SN 1987A, as shown in \citet{kifonidis06}.

\subsection{Aspherical infall}

\begin{figure}[ht]
 \epsscale{0.6}
 \plotone{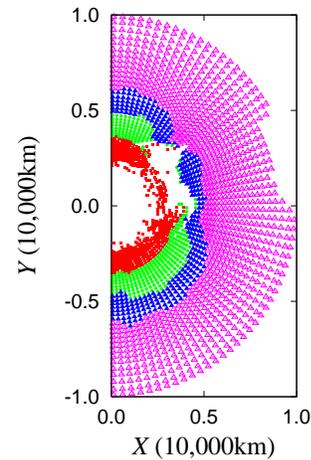}
 \caption{Initial position of SN ejecta that are released from the inner region ($r_{\rm cc}< 10,000\km$)
 for model with $L_{\nu_e} = 4.5 \times 10^{52} \rm \, erg \, s^{-1}$.
 }
\label{fig:inipos}
\epsscale{1.0}
\end{figure}

The value of $M_{\rm ej,in}$ depends on $E_{\rm exp}$ as well as aspherical matter infall to the neutron star,
while the masses of \nuc{Ni}{56} correlates with $E_{\rm exp}$ (Figs.\,\ref{fig:Ep-Lnu}(a) and \ref{fig:mass}).
Figure \ref{fig:inipos} shows the position of SN ejecta before the core collapse 
for model with $L_{\nu_e} = 4.5 \times 10^{52} \rm \, erg \, s^{-1}$.
Filled squares, filled circles, filled triangles, and open triangles indicate ejecta, 
which correspond to the complete Si burning ($T_{\rm max,9} \ge 5$ where $T_{\rm max,9} = T_{\rm max}/10^9 \K$), 
the incomplete Si burning ($4 \le T_{\rm max,9} \le 5$), 
the O burning ($3.3 \le T_{\rm max,9} \le 4$), and the C/Ne burning ($1 \le T_{\rm max,9} \le 3.3$), 
respectively.
We find that whole of the iron core ($\sim 2,000\km$) collapses to the proto neutron star
and that the infall of material to the proto neutron star is aspherical; 
Larger amounts of the gas infall from the upper hemisphere compared with from the lower half.
Larger amounts of material aspherically infall to the neutron star
for model with $L_{\nu_e} = 4.5 \times 10^{52} \rm \, erg \, s^{-1}$
compared with model with $L_{\nu_e} = 5.0 \times 10^{52} \rm \, erg \, s^{-1}$.
Though $E_{\rm exp}$ is comparable, 
$M_{\rm ej,in}$ for $L_{\nu_e} = 4.5 \times 10^{52} \rm \, erg \, s^{-1}$
is smaller than that for $L_{\nu_e} = 5.0 \times 10^{52} \rm \, erg \, s^{-1}$.

\subsection{Aspherical distribution of energy and nuclei}
\label{sec:aspherical_distribution}

\begin{figure*}[ht]
 \epsscale{1.0} 
 \plotone{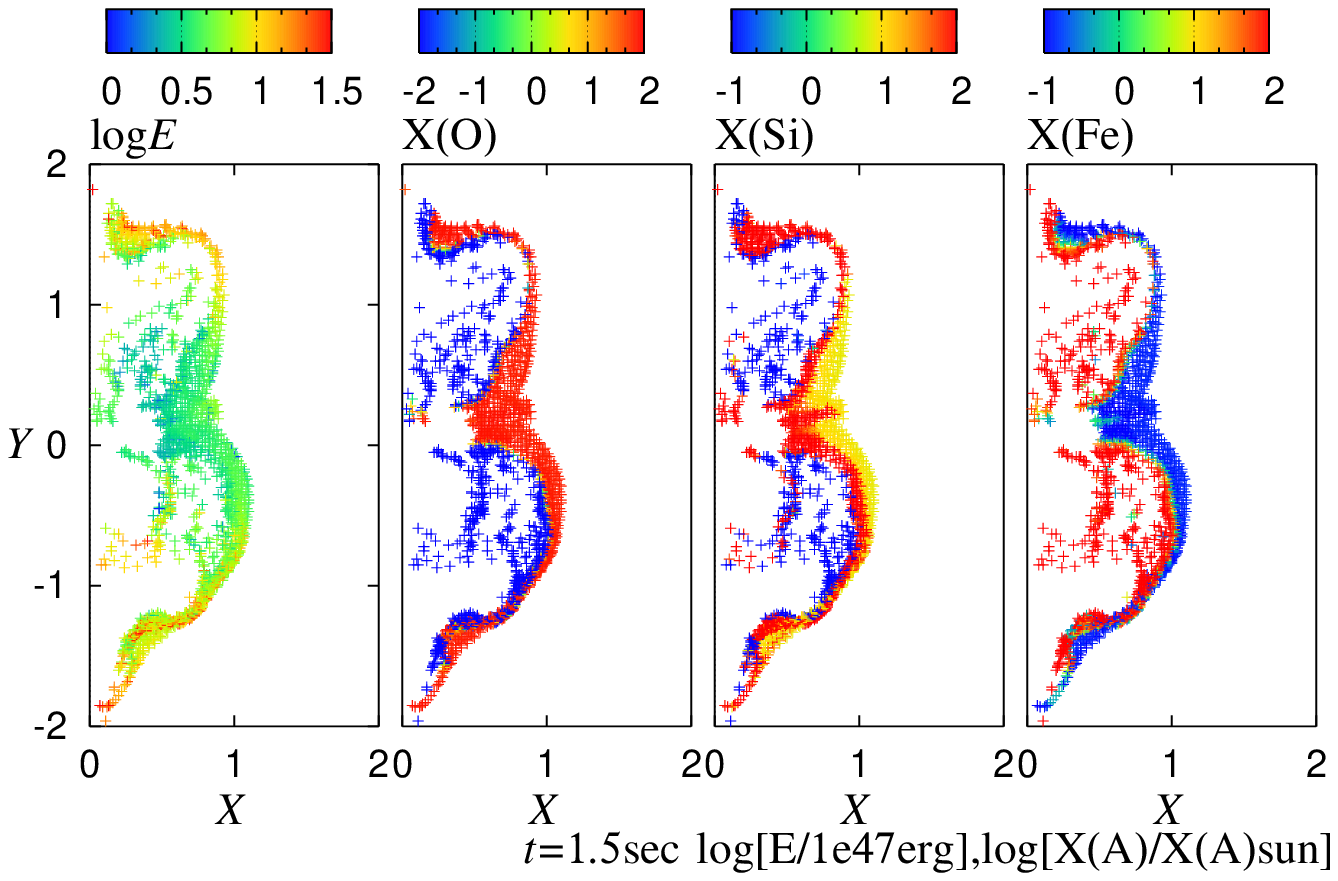}
 \caption{
 Distribution of the energy and abundances of the ejecta
 for $L_{\nu_e} = 4.5 \times 10^{52} \rm \, erg \, s^{-1}$
 at $t = 1.5\s$
 in a region of $0\le x \le 20,000\km$ and $-20,000\km \le y \le 20,000\km$.
 Abundances are shown in units of those in the solar system
 for \nuc{O}{16}, \nuc{Si}{28}, and Fe, which is the sum of \nuc{Fe}{56}, \nuc{Ni}{56}, and \nuc{Fe}{54}.
 }
 \label{fig:dist-ex}
 \epsscale{1.0}
\end{figure*}
Distributions of the energy and abundances of SN ejecta are highly aspherical, 
as one can expect from the entropy distribution (Fig.\,\ref{fig:entropy-hE}).
The distributions are shown for $L_{\nu_e} = 4.5 \times 10^{52} \rm \, erg \, s^{-1}$ at $t = 1.5\s$
in Figure \ref{fig:dist-ex}.
The energy and abundances are presented in units of $10^{47}\rm erg$ and normalized by the solar abundances, 
respectively, for logarithm scale.  
It should be emphasized that the distributions change during later expansion phase~\citep{kifonidis06,gawryszczak10}.
We note that the secondary ejecta that is located in outer layers ($r_{\rm ej,cc} > 10,000\km$) 
are not shown in the figures. 
The high energy ejecta concentrates on the shock front, in particular 
in polar regions ($\theta < \pi/4, \theta > 3\pi/4$).
Both Si and O burning proceed to produce \nuc{Ni}{56} and \nuc{Si}{28} abundantly in the region.
We emphasize that a deformed shell structure forms with Fe, Si and O layers from inside to outside 
in the secondary ejecta heated via the shock wave.
Composition of the secondary ejecta depends chiefly on $T_{\rm max}$, which is determined via the shock heating.
The shell structure is therefore formed in the secondary ejecta, because $T_{\rm max}$ gradually decreases 
from inner to outer layers, where the structure is highly aspherical and deformed.  
We note that Ca is co-produced with \nuc{Si}{28}, as in the spherical models.

The primary ejecta however do not form such a structure.
This is because physical properties of the ejecta, which are mainly heated via neutrinos to be 
$>10^{10}\K$ (Fig. \ref{fig:phys-evol-pe}), are independent from their positions,
and thus their compositions are also independent.
A fraction of the primary ejecta that are mainly composed of Fe and Ni is found to be mixed into the deformed layers, 
while a small amount of the secondary ejecta falls towards the neutron star through near the equatorial plane.
Consequently, the ejecta with abundant O and/or Si appear in the central region ($r < 2,000\km$).
Due to such mixings and infall, some of the Si- and O-rich ejecta can 
penetrate deeper into the Fe-rich ejecta, which is 
in sharp contrast to spherical models.

\begin{figure}[ht]
 \plotone{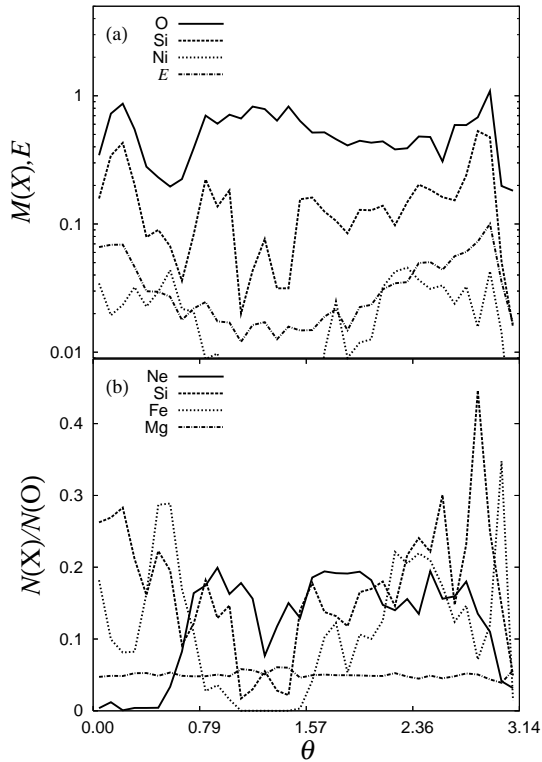}
 \caption{
 (a) Angular distribution of the averaged energy and masses of nuclei 
 of the inner ejecta ($r_{\rm ej,cc} \le 10,000\km$) over a radial direction 
 for $L_{\nu_e} = 4.5 \times 10^{52} \rm \, erg \, s^{-1}$ at $t = 1.5\s$.
 Solid, dashed, dotted, and dash-dotted lines indicate
 the averaged mass of \nuc{O}{16}, \nuc{Si}{28}, and \nuc{Ni}{56} and the averaged energy, 
 respectively. 
 The masses and energy are shown in units of $0.01\Ms$ and $10^{51}\rm erg$, respectively.
 (b) Angular distribution of number ratios of nuclei to \nuc{O}{16} of the inner ejecta ($r_{\rm ej,cc} \le 10,000\km$)
 averaged over a radial direction 
 for $L_{\nu_e} = 4.5 \times 10^{52} \rm \, erg \, s^{-1}$ at $t = 1.5\s$.
 Solid, dashed, dotted, and dash-dotted lines indicate
 the ratios of \nuc{Ne}{20}, \nuc{Si}{28}, iron, and \nuc{Mg}{24}, respectively. 
 The iron consists from \nuc{Ni}{56}, \nuc{Fe}{56}, and \nuc{Fe}{54}.
 }
 \label{fig:asymmetry}
 \epsscale{1.0}
\end{figure}

We have averaged energy and masses of nuclei of the inner ejecta over a radial direction. 
We note that the energy and masses of the outer ejecta ($r_{\rm ej,cc} > 10,000\km$) are not included 
in an averaging procedure.
If we include ejecta from the outer layers, the asymmetry of \nuc{O}{16} becomes small, 
because of the existence of spherically distributed \nuc{O}{16} in the layers.
Figure \ref{fig:asymmetry}(a) shows the averaged energy and masses of \nuc{O}{16}, \nuc{Si}{28}, and \nuc{Ni}{56}.
Asphericity of the energy is prominent;
Ratio of the energy at $\theta=\pi/2$ to that at $\theta=0$ is $\sim 4$ and 
ratio of the maximum to the minimum energy is $\sim 8$.
We note that the ratios are comparable to those in two-dimension models~\citep[models A2 and A3]{nagataki97}.
Only a small amount of \nuc{Ni}{56} exists near the equatorial plane ($\theta \sim (0.8-1.6) \,\rm rad.$), 
where the energy is lower compared with that near the polar axis.
The asymmetry of \nuc{Si}{28} mass is similar to that of the energy and smaller than that of \nuc{Ni}{56}.
Mass distribution of \nuc{Ti}{44} is very similar to that of \nuc{Ni}{56}, as in spherical models, 
because both nuclei are produced in the secondary ejecta through $\alpha$-rich freezeout.
The peaks of \nuc{Si}{28} mass around the polar directions are misaligned 
with those of \nuc{Ni}{56} around $\theta \sim \pi/4$ and $3\pi/4$, 
although the misalignment could disappear due to the lateral motion of the ejecta,
during later explosion phase ($\le 300\s$) as pointed out in \citet{gawryszczak10}.
We note that the lateral motion does not follow during the later phase in the present study.

In addition to the mass fractions of nuclei, 
number ratios of nuclei relative to O (\nuc{O}{16}) are important quantities, 
because the ratios are indicative to the production and destruction mechanism of nuclei.
Angular distributions of the number ratio are shown 
for the inner ejecta ($r_{\rm ej,cc} \le 10,000\km$) averaged over a radial direction
for $L_{\nu_e} = 4.5 \times 10^{52} \rm \, erg \, s^{-1}$ in Figure \ref{fig:asymmetry}(b).
Ratio of Ne (\nuc{Ne}{20}, solid line) is 0.1-0.2 for a direction where the O/Ne burning does not effectively operate,
while it becomes much lower in a region where the O/Ne burning effectively operates to synthesize
abundant Si and/or Fe relative to O.
Distribution of the number ratio $N({\rm Mg})/N({\rm O})$ (\nuc{Mg}{24}, dash-dotted line) 
is similar to the spherical case, 
because Mg is produced during both the stellar evolution and the explosion through the same process,
or the O/Ne burning, with which $N({\rm Mg})/N({\rm O})$ becomes $\sim 0.05$ for a $15\Ms$ progenitor.
Here $N({\rm X})$ is the number of nuclei, X.
Ratio of Si (\nuc{Si}{28}, dashed line) anti-correlates with that of Fe 
(the sum of \nuc{Ni}{56}, \nuc{Fe}{56}, and \nuc{Fe}{54}, dotted line).
This is because Fe is produced via the burning of Si.

\section{Discussions}
\label{sec:discussion}

\subsection{Dependences on neutrino temperatures}
\label{sec:neutrino-temperature}

\begin{deluxetable}{ccccc ccccc c}
\tabletypesize{\scriptsize}
\tablecaption{
 Model parameters and properties of the explosion for nine exploded models\label{tab:model}
 }
\tablehead{%
\colhead{$T_{\nue}$}
 & \colhead{$T_{\nueb}$}
 & \colhead{$L_{\nue}$}
 & \colhead{$E_{\rm exp}$}
 & \colhead{$t_{\rm exp}$}
 & \colhead{$M_{\rm pNS}$}
 & \colhead{$M_{\rm ej,in}$}
 & \colhead{$M$(\nuc{Ni}{56})}
 & \colhead{$M$(\nuc{Ti}{44})}
 }
\startdata
  4.0 &   5.0 &   3.9 &   0.45 &   1.20 &   1.84 &   0.20 & 1.85e-2 & 5.29e-6 \\ 
  4.0 &   5.0 &   4.0 &   0.80 &   0.31 &   1.70 &   0.34 & 2.72e-2 & 1.31e-5 \\ 
  4.0 &   5.0 &   4.2 &   1.02 &   0.31 &   1.65 &   0.39 & 4.51e-2 & 8.04e-6 \\ 
  4.0 &   5.0 &   4.5 &   1.30 &   0.23 &   1.62 &   0.41 & 5.49e-2 & 1.52e-5 \\ 
  4.0 &   5.0 &   4.7 &   1.03 &   0.21 &   1.54 &   0.49 & 5.10e-2 & 1.14e-5 \\ 
  4.0 &   5.0 &   5.0 &   1.30 &   0.18 &   1.56 &   0.48 & 5.44e-2 & 1.05e-5 \\
  3.6 &   4.5 &   4.0 &   0.35 &   1.41 &   1.93 &   0.11 & 1.10e-2 & 6.95e-6 \\ 
  3.6 &   4.5 &   4.5 &   0.35 &   1.19 &   1.87 &   0.17 & 1.40e-2 & 5.53e-6 \\ 
  3.6 &   4.5 &   5.0 &   0.45 &   0.62 &   1.83 &   0.21 & 1.73e-2 & 6.38e-6 
\enddata
\tablecomments{%
 Each column shows $T_{\nue}$, $T_{\nueb}$, $L_{\nue}$, $E_{\rm exp}$, $t_{\rm exp}$ and masses
 ($M_{\rm pNS}$, $M_{\rm ej,in}$, $M$(\nuc{Ni}{56}), and $M$(\nuc{Ti}{44})),
 in units of {\rm MeV}, {\rm MeV}, $10^{52} \erg \s^{-1}$, $10^{51} \erg$, $\s$, and $M_{\odot}$, respectively.
}
\end{deluxetable}

\begin{figure}[ht]
 \plotone{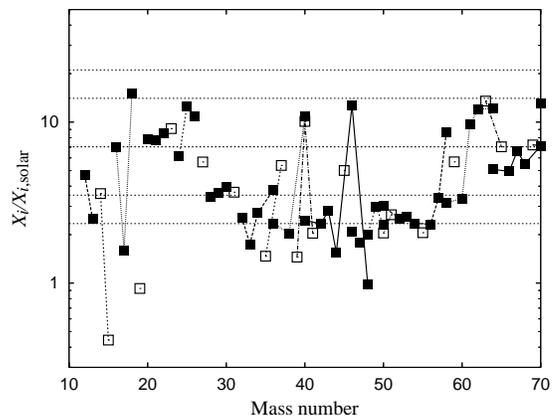}
 \caption{ Same as Figure \ref{fig:opf} but 
 for model with low neutrino temperatures and  $L_{\nu_e} = 5.0 \times 10^{52} \rm \, erg \, s^{-1}$.
 The factors are very similar to those for a model with high neutrino temperatures
 and  $L_{\nu_e} = 3.9 \times 10^{52} \rm \, erg \, s^{-1}$,
 as shown in Figure \ref{fig:opf}(b).
 }
 \label{fig:opf-l50}
 \epsscale{1.0}
\end{figure}

We have assumed the neutrino spheres
with given luminosities and with the Fermi-Dirac distribution of neutrino temperatures,
and considered models with neutrino temperatures, 
$T_{\nu_e}$, $T_{\bar{\nu_e}}$, and $T_{\nu_x}$ 
as $4 \, \rm MeV$ and $5 \, \rm MeV$, and $10 \, \rm MeV$, respectively, 
The temperatures are however slightly lower and change in time,
as shown in simulations using a more elaborate numerical code~\citep{marek09a,marek09b}.
In order to evaluate dependences of hydrodynamic and nucleosynthetic results on
neutrino temperatures, 
we have performed simulations for three models with lower neutrino temperatures;
$T_{\nu_e} = 3.6 \, \rm MeV$, $T_{\bar{\nu_e}}= 4.5 \, \rm MeV$, and $T_{\nu_x} = 9 \, \rm MeV$,
and $L_{\nu_e} = $ (4.0, 4.5, and 5.0) $\times 10^{52} \rm \, erg \, s^{-1}$.
We find that, for a given $L_{\nu_e}$, the explosions are much weaker for a lower $T_\nu$ model, 
in spite of only 10\% changes in neutrino temparatures.
Table 1 summarizes hydrodynamic and nucleosynthetic properties 
for the nine models that produce explosions,
in which $T_{\nu}$s are changed in the three ways as mentioned above.
The explosion energies are (0.35, 0.35, and 0.45) $\times 10^{51} \erg$
for models with $L_{\nu_e} = $ (4.0, 4.5, and 5.0) $\times 10^{52} \rm \, erg \, s^{-1}$, respectively.
Mass of the ejecta from the inner region ($r_{\rm ej,cc} \le 10,000\km$), $M_{\rm ej,in}$, 
and \nuc{Ni}{56} mass are $0.11-0.21\Ms$ and $0.011-0.017\Ms$, respectively.
Figure \ref{fig:opf-l50} shows overproduction factors
for the low $T_\nu$ model with $L_{\nu_e} = 5.0 \times 10^{52} \rm \, erg \, s^{-1}$.
We emphasize that the overproduction factors are very similar to those 
for the high $T_\nu$ model with $L_{\nu_e} = 3.9 \times 10^{52} \rm \, erg \, s^{-1}$,
in which the explosion energy and the ejecta mass from the inner region 
are comparable to the above model, or $0.45 \times 10^{51} \erg$ and $0.020\Ms$, respectively.
We conclude that the abundances of the ejecta do not directly depend on $T_\nu$ 
but mainly depend on the explosion energy and the ejecta mass.

\subsection{Effects of the $\nu$ interactions on heavy nuclei}
\label{sec:neutrino-effects}

\begin{figure}[ht]
 \plotone{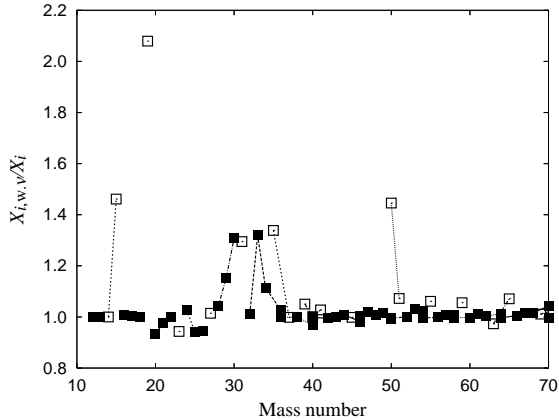}
 \caption{Ratios of the abundances of SN ejecta taking into account 
 $\nu$ interactions on heavy nuclei to those
 without the $\nu$ interactions, as a function of $A$
 for $L_{\nu_e} = 4.5 \times 10^{52} \rm \, erg \, s^{-1}$.
 }
 \label{fig:comp-neutrino}
 \epsscale{1.0}
\end{figure}

Effects of neutrino interactions on SN ejecta have been investigated for spherical SN explosion~\citep{whhh90,ww95}.
We have calculated abundances of the ejecta taking into account $\nu$ interactions on heavy nuclei~\citep{goriely01}
for $L_{\nu_e} = 4.5 \times 10^{52} \rm \, erg \, s^{-1}$.
Cross sections for the neutrino reactions are taken from \citet{whhh90}, 
in which rates are included for neutral and charged current reactions on He and from C to Kr.
Figure \ref{fig:comp-neutrino} shows ratios of the abundances of the ejecta with the $\nu$ interactions
to those without the interactions.
We find that the increase of the abundances due to the interactions of $\nu$ on heavy nuclei 
are up to 50\%, except for \nuc{F}{19}, whose ratio is 2.1.
\nuc{F}{19} is enhanced via the neutral and charged current interactions of \nuc{Ne}{20}, 
or \nuc{Ne}{20}$(\nu\nu^{'},\rm p)$\nuc{F}{19} and \nuc{Ne}{20}$(\nueb,e^+\rm n)$\nuc{F}{19}, respectively.
\nuc{N}{15} and \nuc{V}{50} are synthesized through 
\nuc{O}{16}$(\nu\nu^{'},\rm p)$\nuc{N}{15} and \nuc{Cr}{50}$(\nueb,e^+)$\nuc{V}{50}, respectively.
We note that the enhancement of light elements, such as Li, Be, and B, via the $\nu$ interactions
is not fully taken into account in our calculation, because the composition of ejecta from outer layers
($r_{\rm ej,cc} >10,000\km$) is fixed to be the pre-SN composition,
although the production of these elements is efficient in the outer layers 
through neutrino interactions~\citep{whhh90, yoshida08}.
Moreover, neutrino absorptions on D may be important for nucleosynthesis as well as the dynamics of
the SN explosion~\citep{nakamura09}.

\subsection{Dependences of abundances on nuclear reaction rates}
\label{sec:reaction-rates}

\begin{figure}[ht]
 \plotone{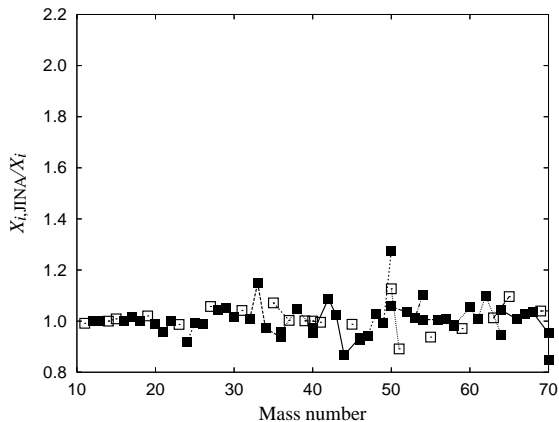}
 \caption{Ratios of the abundances of SN ejecta using reaction rates 
 in the JINA REACLIB V1.0 database to those in REACLIB database, as a function of $A$
 for $L_{\nu_e} = 4.5 \times 10^{52} \rm \, erg \, s^{-1}$.
 }
 \label{fig:comp-rate}
 \epsscale{1.0}
 \end{figure}

We have adopted reaction rates mainly taken from the REACLIB database 
in our nuclear reaction network presented in \S \ref{sec:network}.
The database is recently updated and continuously maintained by the Joint Institute for 
Nuclear Astrophysics (JINA) REACLIB project~\citep{cyburt10}.
We have calculated abundances of the ejecta for $L_{\nu_e} = 4.5 \times 10^{52} \rm \, erg \, s^{-1}$
adopting reaction rates taken from the JINA REACLIB V1.0 database.
Figure \ref{fig:comp-rate} shows ratios of the abundances of the ejecta adopting reaction rates 
in the JINA REACLIB V1.0 database to those in REACLIB database. 
We find that the differences are small, up to 30\%, 
between abundances with the JINA REACLIB V1.0 and REACLIB databases.
If we use newly evaluated $\alpha$-capture rates on \nuc{Ca}{40} and \nuc{Ti}{44}, 
yields of \nuc{Ti}{44} are likely to be lower~\citep{hoffman10}.

\subsection{Uncertainty of abundances of the ejecta}
\label{sec:uncertainty}

\begin{figure}[ht]
 \plotone{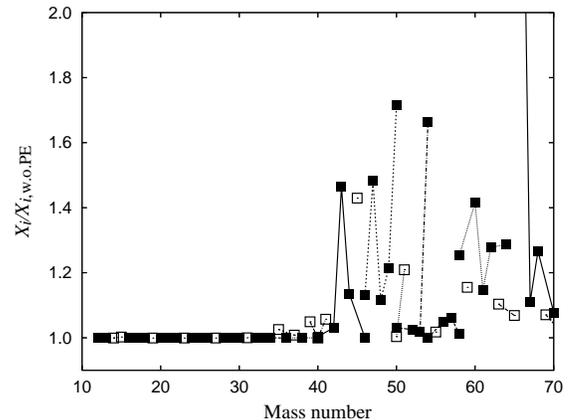}
 \caption{Ratios of the abundances of SN ejecta to those 
 without the primary ejecta, as a function of $A$
 for $L_{\nu_e} = 4.5 \times 10^{52} \rm \, erg \, s^{-1}$.
 }
 \label{fig:comp-w.o.pe}
 \epsscale{1.0}
\end{figure}

As shown in \S \ref{sec:pe-se}, 
the ejecta consists of the primary ejecta, whose $Y_e$ and thus abundances are highly uncertain
and the secondary ejecta, which have a relatively definite composition and are much heavier than the primary ejecta.
In order to clarify uncertainty on the estimate of abundances of the ejecta,
we have evaluated abundances of the ejecta being integrated over ejecta without the primary ejecta and
have compared above the abundances with those summed over all the ejecta (Fig.\,\ref{fig:opf}(a)).
Figure \ref{fig:comp-w.o.pe} shows ratios of the abundances of all the ejecta to those without the primary ejecta
for $L_{\nu_e} = 4.5 \times 10^{52} \rm \, erg \, s^{-1}$.
We find that 
abundances of most of nuclei with $A \le 70$ do not largely change within a factor of 2.
The overproduction factors are therefore very similar to those in Figure \ref{fig:opf}(a).
The ratios are greater than 1.5 for \nuc{Ca}{43}, \nuc{Sc}{45}, \nuc{Ti}{47}, \nuc{Ti}{50}, 
\nuc{Cr}{54}, \nuc{Ni}{60}, \nuc{Zn}{64}, \nuc{Zn}{66}, and \nuc{Ge}{70}, in particular, 
9.0, 2.8, and 3.1 
for \nuc{Zn}{64}, \nuc{Zn}{66}, and \nuc{Ge}{70}, respectively.
These nuclei are therefore chiefly synthesized in the primary ejecta, 
not in the secondary ejecta.
Moreover, 
\nuc{Ti}{50}, \nuc{Cr}{54}, \nuc{Ni}{60}, \nuc{Zn}{64}, \nuc{Zn}{66}, and \nuc{Ge}{70} are 
synthesized in neutron-rich primary ejecta, 
while \nuc{Ca}{43}, \nuc{Sc}{45}, \nuc{Ti}{47}, and \nuc{Ni}{60} are abundantly produced in 
proton-rich primary ejecta through $\nu$p-processes~\citep{frohlich06a,frohlich06b,pruet06,wanajo06}.

In short, 
abundances of \nuc{Zn}{64}, \nuc{Zn}{66}, and \nuc{Ge}{70} are highly uncertain 
but those of the other nuclei are relatively definite.
If a fraction of the primary ejecta could become much larger, abundances might be highly 
uncertain, in particular for the nuclei that are produced in the primary ejecta.
The fraction is however unlikely to be much larger, because the ejection of the secondary
ejecta is driven by the shock wave caused by the primary ejecta.
It should be noted that the mass fractions of the primary ejecta to the secondary ejecta 
are comparable for all the models, 
although the explosion energies and ejected masses from the inner region diverse among the models.

Moreover, matter near the proto neutron star is blown off via neutrino-driven winds 
during later evolution of the star ($>2\s$).
The ejecta could be not neutron-rich but proton-rich ($Y_e \sim 0.5 -0.6$) 
and have high entropy ($> 50 k_{\rm B}$)~\citep{fischer10, hudepohl10}.
The $\nu$p-process could operate in the ejecta to synthesize 
light p-nuclei~\citep{frohlich06a,frohlich06b,pruet06,wanajo06}.
However, the process may have not large contributions to the ejected masses of abundant nuclei, 
since the mass ejection rate is small through the winds at the later epoch~\citep{hudepohl10}.

\subsection{Comparison with observations on abundances of SN remnants}

Recently, \citet{kimura09,uchida09} have analyzed the metal distribution
of the Cygnus loop by using the data obtained by the Suzaku and XMM-Newton observations.
It is pointed out that the progenitor of the Cygnus loop is a 
core-collapse supernova explosion whose progenitor mass ranges in $\sim 12-15\Ms$.
The ejecta distributions are asymmetric to the geometric center: the ejecta of 
O, Ne are distributed more in the north-west rim, while the ejecta of Si and Fe are
distributed more in the south-west of the Cygnus loop. Since the material in this 
middle-aged supernova remnant has not been completely mixed yet, the observed 
asymmetry is considered to still remain a trace of inhomogeneity 
produced at the moment of explosion. These evidences may allow us to speculate 
that the asymmetry of the heavy element observed in the Cygnus loop may come from globally 
asymmetric explosions explored in this work. We try to seek the relevance in the following.

\begin{figure}[ht]
 \plotone{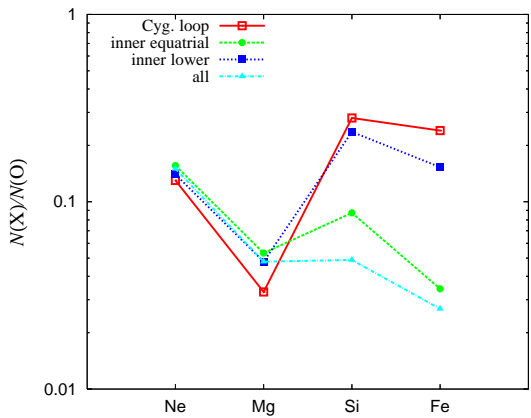}
 \caption{
 Number ratios of Ne, Mg, Si, and Fe relative to O observed in Cygnus loop
 and those of ejecta 
 for $L_{\nu_e} = 4.5 \times 10^{52} \rm \, erg \, s^{-1}$ at $t = 1.5\s$.
 Solid line indicates ratios observed in Cygnus loop, while
 dotted, dashed, and dash-dotted line indicate  ratios averaged over 
 ejecta from an inner equatorial region ($r_{\rm ej,cc} \le 10,000\km$ and $\pi/3 < \theta < 2\pi/3$), 
 ejecta from an inner polar region ($r_{\rm ej,cc} \le 10,000\km$ and $\theta \ge 2\pi/3$), 
 and all the ejecta, respectively.
 }
 \label{fig:oratio}
 \epsscale{1.0}
\end{figure}

Figure \ref{fig:oratio} shows 
number ratios of Ne, Mg, Si, and Fe relative to O observed in Cygnus loop and those of ejecta 
for $L_{\nu_e} = 4.5 \times 10^{52} \rm \, erg \, s^{-1}$.
We find that the ratios observed in Cygnus loop (solid line) are comparable to those averaged over 
ejecta from the inner lower region ($r_{\rm ej,cc} \le 10,000\km$ and $\theta \ge 2\pi/3$) (dotted line).
The ratios averaged over all the ejecta (dash-dotted line) are same as those averaged over 
ejecta from the inner equatorial region ($r_{\rm ej,cc} \le 10,000\km$ and $\pi/3 < \theta < 2\pi/3$) (dashed line).
We note that the ratios averaged over 
the ejecta from the inner upper region ($r_{\rm ej,cc} \le 10,000\km$ and $\theta \le \pi/3$)
are comparable to but slightly lower than those averaged over the ejecta from the inner lower region (dotted line).

$N({\rm Ne})/N({\rm O})$ and $N({\rm Mg})/N({\rm O})$ are independent from an averaging region
and are comparable to all our aspherical models.
These ratios are therefore concluded to be determined during the hydrostatic evolution of the progenitor 
and thus are a good indicator for the progenitor mass. 
On the other hand, ratios of Si and Fe relative to O are much higher in the inner region, 
in particular in the inner lower region, compared with the outer region ($\ge 10,000\km$).
$N({\rm Si})/N({\rm O})$ and $N({\rm Fe})/N({\rm O})$ averaged over all the ejecta 
are much smaller than those observed in Cygnus loop.
Therefore incomplete radial mixing of Fe and Si with O is required 
to be explained the high Fe and Si ratios in Cygnus loop.
In fact, the averaged abundances of the Cygnus loop have a correlation
with the Si- and Fe- rich regions \citep{kimura09,uchida09}.
Our results suggest that the abundance ratios for Ne and Mg as well as Si and Fe 
observed in the Cygnus loop could be well reproduced with the SN ejecta from the inner region
of a $15\Ms$ progenitor.

\subsection{Three dimensional effects}

In order to draw a robust conclusion to the findings obtained in the current
2D simulations, it is indispensable to move on to three dimensional (3D) simulations
(e.g.,\citet{blondin07a,iwakami08,iwakami09,nordhaus10,wongwathanarat10}).
In 2D, the growth of SASI and the large-scale convection tend to develop
along the coordinate symmetry axis preferentially, 
thus suppressing the anisotropies in explosions as well as in the resulting 
explosive nucleosynthesis.
An encouraging news to us is that 3D simulations cited above are, at least, 
in favour of the SASI-aided low-modes explosions, in which the major axis of the 
expanding shock has a certain direction at the moment of explosion
(albeit free from the coordinate symmetry axis). 
By incorporating the present scheme to the 3D simulations of \citet{iwakami08,iwakami09},
we plan to clarify these issues as a sequel of this study. 

\section{Summary}
\label{sec:summary}

We have investigated the explosive nucleosynthesis in
the delayed neutrino-driven, aspherical SN explosion aided by SASI,
based on 2D, axisymmetric hydrodynamic simulations of 
the explosion of a non-rotating 15$M_\odot$ star.
We employed a hydrodynamic code with a simplified light-bulb neutrino transport scheme.
We have approximately taken into account neutrino heating and cooling as well as 
the evolution of electron fraction due to weak interactions,
both in the hydrodynamic simulations and nucleosynthetic calculations.
Neutrinos are assumed to be isotropically emitted from the neutrino spheres
with given luminosities and with the Fermi-Dirac distribution of given temperatures.
We have performed simulations with the temperatures and luminosities of $\nu_e$, $\bar{\nu}_e$ and $\nu_x$
constant in time. 
We have followed abundance evolution of SN ejecta using the nuclear reaction network coupled with
an evolution equation of the electron fraction of the ejecta.

We summarize our results as follows, 
\begin{enumerate}
 \item The stalled shock revives due to the neutrino heating aided by SASI
       for cases with $L_{\nu_e} \ge 3.9 \times 10^{52} \rm erg \, s^{-1}$
       and the aspherical shock passes through the outer layers of the star 
       ($\ge \rm 10,000 \, km$).
       Evaluated explosion energies roughly correlate with neutrino luminosities.
       For models with larger luminosities, the explosion occurs earlier and mass of a neutron star becomes lighter.
 \item Whole of the iron core of the progenitor collapses to the proto neutron star.
       The infall of material to the star is aspherical.
       Larger amounts of the gas infall from directions with lower explosion energies.
 \item Abundances of the neutrino-heated ejecta are highly uncertain, in particular for neutron-rich ones, 
       due to the uncertainty in the estimate on $Y_e$.
       On the other hand, the shock-heated ejecta has definite abundances, 
       which depend on the maximum temperature mainly.
       The uncertainty in the estimate of the masses and abundances of abundant nuclei in the SN ejecta 
       is small because of the small fraction of the neutrino-heated ejecta.
 \item Abundance pattern of the supernova ejecta is similar to 
       that of the solar system, for cases with the mass of the ejecta from the inner region 
       ($\le 10,000\km$), $M_{\rm ej,in} = (0.4-0.5)\Ms$, which corresponds to models with
       a high explosion energy of $ \simeq 10^{51} \rm erg$.
       Masses of a neutron star remnant, estimated to be $(1.54-1.62) \Ms$ for these models, 
       are comparable to the baryonic mass of the neutron star observed in neutron-star binaries.
       $E_{\rm exp}/M_{\rm ej}$ evaluated for the models ($(0.92 - 1.2) \times 10^{50} \erg/\Ms$)
       are comparable to the estimate in SN 1987A.
 \item Underproduction of \nuc{Ti}{44} and overproduction of \nuc{Ni}{62}, which appear
       in spherical models, are also shown in our 2D calculations.
       The overproduction of \nuc{Ni}{62} possibly inherits the uncertainty
       not only in the progenitor model but also of the change in $Y_e$ during the SN explosion.
       On the other hand, \nuc{Zn}{64}, which is underproduced in a spherical model, 
       is found to be abundantly produced in our 2D model, 
       although the abundance and mass are uncertain.
 \item Distributions of nuclei and energy are highly aspherical in the SN ejecta, 
       although the progenitor is non-rotating and has spherical symmetric configuration.
       The shock-heated ejecta forms aspherical and deformed shell-like structure composed of Fe, Si, and O 
       from inside to outside.
       The neutrino-heated ejecta do not have any definite structure and fractions of the ejecta are mixed into
       the shell-like structure of the shock-heated ejecta.
 \item The asymmetry of \nuc{Si}{28} mass distribution 
       is similar to that of the explosion energy and smaller than that of \nuc{Ni}{56}.
       \nuc{Ca}{40} and \nuc{Ti}{44} are accompanied with \nuc{Si}{28} and \nuc{Ni}{56}, respectively.
 \item The ratios for Ne, Mg, Si and Fe observed in Cygnus loop 
       are well reproduced with the SN ejecta from the inner region of the $15\Ms$ progenitor.
       The number ratios for Ne and Mg relative to O seem to be good indicators for the mass of a progenitor
       of a core-collapse SN.
\end{enumerate}

\acknowledgments{
S.F. is grateful to K. Sumiyoshi for fruitful discussions.
K.K. thanks to S.Nagataki for stimulating discussions and also 
thanks to K. Sato and S. Yamada for continuing encouragements.
This work is partly supported by a grant for Basic Science Research Projects 
from the Sumitomo Foundation (No. 080933) and
Grant-in-Aid for Scientific Research from the Ministry of
Education, Culture, Sports, Science and Technology of Japan
(Nos. 19540309, 20740150, 22540297). 
}


\newcommand{\ye}{Y_{\rm e}} 
\newcommand{\Ye}{Y_{\rm e}} 
\newcommand{\yl}{Y_{\rm l}} 
\newcommand{\Yl}{Y_{\rm l}} 
\newcommand{\nube}{{\bar\nu_e}} 
\newcommand{\nub}{\bar \nu} 
\newcommand{\nubar}{\bar \nu} 
\newcommand{\nuxbar}{\bar \nu_x} 
\newcommand{\nuebar}{{\bar \nu}_{\rm e}} 
\newcommand{\num}{\nu_\mu} 
\newcommand{\numbar}{{\bar \nu}_\mu} 
\newcommand{\nut}{\nu_\tau} 
\newcommand{\nutbar}{{\bar \nu}_\tau} 
\newcommand{\Msol}{M_{\odot}}
\newcommand{\epsnu}{\epsilon_{\nu}}
\newcommand{\epsnui}{\epsilon_{\nu_i}}
\newcommand{\epsnb}{\epsilon_{\nubar}}
\newcommand{\epsnubi}{\epsilon_{\nubar_i}}
\newcommand{\epsne}{\epsilon_{\nue}}
\newcommand{\epsna}{\epsilon_{\nuebar}}
\newcommand{\epse}{\epsilon_{\rm e}}
\newcommand{\epsem}{\epsilon_{{\rm e}^-}}
\newcommand{\epsep}{\epsilon_{{\rm e}^+}}
\newcommand{\epsepm}{\epsilon_{{\rm e}^{\pm}}}

\newcommand{\eavnu}{\langle\epsnu\rangle}
\newcommand{\eavnb}{\langle\epsnb\rangle}
\newcommand{\eavne}{\langle\epsne\rangle}
\newcommand{\eavna}{\langle\epsna\rangle}
\newcommand{\eave}{\langle\epse\rangle}
\newcommand{\eavep}{\langle\epsep\rangle}
\newcommand{\eavem}{\langle\epsem\rangle}
\newcommand{\eavepm}{\langle\epsepm\rangle}

\newcommand{\eavnut}[1]{\langle\epsnu^#1\rangle}
\newcommand{\eavnbt}[1]{\langle\epsnb^#1\rangle}
\newcommand{\eavnet}[1]{\langle\epsne^#1\rangle}
\newcommand{\eavnat}[1]{\langle\epsna^#1\rangle}
\newcommand{\eavet}[1]{\langle\epse^#1\rangle}
\newcommand{\eavept}[1]{\langle\epsep^#1\rangle}
\newcommand{\eavemt}[1]{\langle\epsem^#1\rangle}
\newcommand{\eavepmt}[1]{\langle\epsepm^#1\rangle}

\newcommand{\eavnus}{\langle\epsnu\rangle^{\!\star}}
\newcommand{\eavnbs}{\langle\epsnb\rangle^{\!\star}}
\newcommand{\eavnes}{\langle\epsne\rangle^{\!\star}}
\newcommand{\eavnas}{\langle\epsna\rangle^{\!\star}}
\newcommand{\eaveps}{\langle\epsep\rangle^{\!\star}}
\newcommand{\eavems}{\langle\epsem\rangle^{\!\star}}

\newcommand{\eavnust}[1]{\langle\epsnu^#1\rangle^{\!\star}}
\newcommand{\eavnbst}[1]{\langle\epsnb^#1\rangle^{\!\star}}
\newcommand{\eavnest}[1]{\langle\epsne^#1\rangle^{\!\star}}
\newcommand{\eavnast}[1]{\langle\epsna^#1\rangle^{\!\star}}
\newcommand{\eavepst}[1]{\langle\epsep^#1\rangle^{\!\star}}
\newcommand{\eavemst}[1]{\langle\epsem^#1\rangle^{\!\star}}
\newcommand{\fnuvac}{f_{\nu,{\rm vac}}}

\newcommand{\Rae}{{\cal R}^{\rm a}_{\nue}}
\newcommand{\Ree}{{\cal R}^{\rm e}_{\nue}}
\newcommand{\Raa}{{\cal R}^{\rm a}_{\nueb}}
\newcommand{\Rea}{{\cal R}^{\rm e}_{\nueb}}
\newcommand{\Qae}{{\cal Q}^{\rm a}_{\nue}}
\newcommand{\Qee}{{\cal Q}^{\rm e}_{\nue}}
\newcommand{\Qaa}{{\cal Q}^{\rm a}_{\nueb}}
\newcommand{\Qea}{{\cal Q}^{\rm e}_{\nueb}}

\begin{appendix}

In this appendix, we summarize our treatment on
the rate of change in $Y_e$ and energy per unit volume, $Q_{\rm N}$ and $Q_{\rm E}$, respectively.
We take into account absorption of electron and anti-electron neutrinos 
as well as neutrino emission through electron and positron captures, 
electron-positron pair annihilation, 
nucleon-nucleon bremsstrahlung, and plasmon-decays.
Moreover, we include the heating term in $Q_{\rm E}$ due to the absorption of neutrinos on \nuc{He}{4} and 
the inelastic scatterings on \nuc{He}{4} via neutral currents~\citep{haxton88,ohnishi07}.
We chiefly follow the treatments in Appendix D of \citet{scheck06} and 
in Appendix B of \citet{ruffert96} (and references therein).

\section{the rate of change in specific energy, $Q_{\rm E}$ }

\subsection{neutrino absorption processes}

The heating rate per unit volume through the absorption of $\nue$ on neutrons is described as
\begin{equation}
  \Qae
    = \sigma c \frac{L_{\nue} \, n_{\rm n}}{4\pi r^2 c f_{\nue}}
   \, \frac{ \eavnet{3} +2\Delta \eavnet{2} +\Delta^2 \eavne }{\eavne} \, \Theta(\eavne),
\end{equation}
where $\Delta = (m_n -m_p)c^2$, $n_n$ is the number density of neutrons, and 
$\sigma = 4 G_{\rm F}^2 m_{\rm e}^2 \hbar^2 / \pi c^2$
$= \frac{1}{4}(3\alpha_{\rm w}^2+1) \sigma_0 / (m_{\rm e}c^2)^2$, 
with the Fermi coupling constant $G_{\rm F}$, the reduced planck constant $\hbar$, the electron mass $m_{\rm e}$, 
$\alpha_{\rm w}=1.254$, and $\sigma_0 = 1.76\times 10^{-44}\,{\rm cm}^{2}$.
Here, the $n$-th energy moment of $\nue$, $\langle\epsnui^n\rangle$, is given by
\begin{equation}
 \langle\epsnui^n\rangle =
 (k_{\rm B} T_{\nu_i})^n \, \frac{ {\cal F}_{2+n}(\eta_{\nu_i}) } { {\cal F}_2(\eta_{\nu_i}) }, \\
\end{equation}
and a factor for the Pauli blocking, $\Theta(\langle\epsnui\rangle)$, is approximated as
\begin{equation}
  \Theta(\langle\epsnui\rangle) =
 1-f_{\rm FD}\left( \frac{\langle\epsnui\rangle+\Delta}{k_{\rm B}T},\eta_{\rm e^-} \right),
\end{equation}
where $\eta_{\nu_i}$ is the chemical potential of $\nu_i$ in units of $k_{\rm B} T$ and set to be 0, 
$\eta_{\rm e^-}$ is the chemical potential of electrons in units of $k_{\rm B} T$, 
$k_{\rm B}$ is the Boltzmann constant,
and ${\cal F}_n(\eta)$ is defined as,
\begin{equation}
 {\cal F}_n(\eta) \equiv
    \int_0^{\infty} \mathrm{d}x  \; x^n \, 
    f_{\rm FD}(x,\eta) 
\end{equation}
using the Fermi-Dirac distribution function, 
\begin{equation}
f_{\rm FD}(x,\eta) = \frac{1}{1 \; + \; \exp(x - \eta)}.
\end{equation}

The heating rate via the absorption of $\nueb$ on protons is given by
\begin{eqnarray}
  \Qaa = \sigma c \frac{L_{\nuebar} \, n_{\rm p}}{4\pi r^2 c f_{\nuebar}} 
   \frac{ \eavnast{3} +3\Delta  \eavnast{2} +3\Delta^2 \eavnas +\Delta^3 \eavnast{0} }{\eavna},
\end{eqnarray}
where the $n$-th energy moment of $\nueb$, $\langle\epsnubi^n\rangle^{\star}$, is given by
\begin{equation}
 \langle\epsnubi^n\rangle^{\star} = (k_{\rm B} T_{\nub_i})^n \, 
  \frac{ {\cal F}_{2+n}(\eta_{\nub_i} - \Delta/k_{\rm B}T_{\nub_i}) }
  { {\cal F}_2(\eta_{\nub_i}) }.
\end{equation}

\subsection{Capture processes}

The energy emission rate of $\nue$ per unit volume 
via electron capture on protons is given by
\begin{equation}
  \Qee = \frac{\sigma c}{2} \; n_{\rm p} \, n_{{\rm e}^-} 
             \left[ \eavemst{3} +2\Delta \eavemst{2} +\Delta^2 \eavems \right],
\end{equation}
and that of $\nueb$ via positron capture on neutrons by
\begin{equation}
  \Qea = \frac{\sigma c}{2}  \; n_{\rm n} \, n_{{\rm e}^+}
             \left[ \eavept{3} +3\Delta \eavept{2} +3\Delta^2 \eavep +\Delta^3 \right].
\end{equation}
Here, the number densities of electrons and positrons are described by
\begin{equation}
 n_{\mathrm{e^\mp}} = {8\pi\over (hc)^3}\,(k_{\mathrm{B}} T)^3 {\cal F}_2(\pm\eta_{\mathrm{e}^-}),
\end{equation}
where $h$ is the Planck constant, 
and the $n$-th energy moments of electrons and positrons, 
$\langle\epse^n\rangle$ and $\langle\epse^n\rangle^{\star}$, are given by
\begin{equation}
 \langle\epse^n\rangle =
 (k_{\rm B}T)^n \, \frac{{\cal F}_{2+n}(\eta_{\mathrm{e}})} {{\cal F}_2(\eta_{\mathrm{e}})}
\end{equation}
and 
\begin{equation}
 \langle\epse^n\rangle^{\star} =
 (k_{\rm B}T)^n \, \frac{{\cal F}_{2+n}(\eta_{\mathrm{e}}-\Delta/k_{\rm B}T)}
 {{\cal F}_2(\eta_{\mathrm{e}})}.
\end{equation}

\subsection{Pair processes}

The energy emission rates of $\nue$ and $\nube$ per unit volume 
via electron-positron pair annihilations are given by
\begin{equation}
 {\cal Q}^{\rm pa}_{\nue} = Q^{\rm pa}_{\nueb} = 
\frac{C_V^2+ C_A^2}{36} 
 \left(\frac{8\pi}{(h c)^3}\right)^2
 \frac{\sigma_0 c}{(m_e c^2)^2} 
 (k_{\rm B}T)^9 [ {\cal F}_{4}(\eta_e) {\cal F}_{3}(-\eta_e) + {\cal F}_{3}(\eta_e) {\cal F}_{4}(-\eta_e) ]
 P_{{\rm pair},\nue} P_{{\rm pair},\nube},
\end{equation}
where $C_A = \frac{1}{2}$, $C_V = \frac{1}{2} + 2 \sin^2 \theta_{\rm W}$, 
and $\sin^2 \theta_{\rm W} = 0.23$.
Here $P_{{\rm pair},\nue}$ and $P_{{\rm pair},\nube}$ are 
factors for the phase space blocking of $\nue$ and $\nube$ for the pair processes, 
respectively and the factor for $\nu_i$ is approximately expressed as
\begin{equation}
P_{{\rm pair},\nu_i} \simeq \left\{ 1 + \exp \left[-\left( 
\frac{1}{2}\frac{{\cal F}_4(\eta_{e})}{{\cal F}_3(\eta_e)} 
+\frac{1}{2}\frac{{\cal F}_4(-\eta_{e})}{{\cal F}_3(-\eta_e)} 
 -\eta_{\nu_i}\right)
\right]\right\}^{-1}.
\end{equation}

The energy emission rate of $\nux$ per unit volume via the pair annihilations is given by
\begin{equation}
 {\cal Q}^{\rm pa}_{\nux} = 
 \frac{(C_V -C_A)^2 +(C_V +C_A -2)^2}{18} 
 \left(\frac{8\pi}{(h c)^3}\right)^2
 \frac{\sigma_0 c}{(m_e c^2)^2} 
 (k_{\rm B}T)^9 [ {\cal F}_{4}(\eta_e) {\cal F}_{3}(-\eta_e) + {\cal F}_{3}(\eta_e) {\cal F}_{4}(-\eta_e) ]
 (P_{{\rm pair},\nux})^2.
\end{equation}

\subsection{Nucleon-nucleon bremsstrahlung processes}

The energy emission rate per unit volume through nucleon-nucleon bremstrahlung of a single neutrino pair 
is well approximated with
\begin{equation}
{\cal Q}^{\rm br}_{\nu\bar{\nu}} = 1.04 \times 10^{30} \zeta 
(X_n^2 +X_p^2 +\frac{28}{3}X_n X_p) \left(\frac{\rho}{10^{14}\gpccm} \right)^2
\left(\frac{T}{1 \mev} \right)^{5.5} \, \erg \, \cm^{-3} \, \s^{-1},
\end{equation}
where $\zeta$ is set to be 0.5~\citep{burrows00}, 
and $X_n$ and $X_p$ are the mass fraction of neutrons and protons, respectively.

\subsection{Plasmon decay processes}

The emission rates of $\nue$ and $\nube$ per unit volume
via plasmon decay are given by
\begin{equation}
R^{\rm pl}_{\nue} = R^{\rm pl}_{\nube} = 
 C_V^2 \frac{\pi^3}{3 \alpha_* (hc)^6} \frac{\sigma_0 c}{(m_e c^2)^2} 
 (k_{\rm B}T)^8 \gamma^6 e^{-\gamma}(1+\gamma)  P_{{\rm plas},\nue} P_{{\rm plas},\nube},
\end{equation}
where $\alpha_* = 1/137.036$ is the fine-structure constant, 
$\gamma = \gamma_0 \sqrt{\eta_e^2 +\pi^2/3}$ with 
$\gamma_0 = 2\sqrt{\alpha_*/3\pi} = 5.565 \times 10^{-2}$, 
$P_{{\rm plas},\nue}$ and $P_{{\rm plas},\nube}$ are 
factors for the phase space blocking of $\nue$ and $\nube$ for 
plasmon decay processes, respectively.
The factor for $\nu_i$ is approximately expressed as
\begin{equation}
P_{{\rm plas},\nu_i} \simeq \left\{ 1 + 
 \exp \left[-\left( 1+ \frac{1}{2}\frac{\gamma^2}{1+\gamma} -\eta_{\nu_i}\right)
\right]\right\}^{-1}.
\end{equation}

The emission rate of $\nux$ per unit volume via plasmon decay is given by
\begin{equation}
R^{\rm pl}_{\nux} = 
(C_V-1)^2 \frac{4\pi^3}{3 \alpha_* (hc)^6}  \frac{\sigma_0 c}{(m_e c^2)^2} 
 (k_{\rm B}T)^8 \gamma^6 e^{-\gamma}(1+\gamma) (P_{{\rm plas},\nux})^2
\end{equation}

The energy emission rates of $\nu_i$ per unit volume
via plasmon decaye is given by
\begin{equation}
 {\cal Q}^{\rm pl}_{\nu_i} =  R^{\rm pl}_{\nu_i} \cdot
k_{\rm B} T \left( 1 + \frac{1}{2}\frac{\gamma^2}{1+\gamma} \right)
\end{equation}

\subsection{Neutrino absorption on helium and inelastic neutrino-helium scatterings}

In addition to the heating processes through the neutrino absorption on nucleons,
the heating processes due to neutrino-helium interactions are taken into account,
as in \citet{ohnishi07}.
Through the absorption of $\nue$ and $\nueb$ and 
the neutrino-helium inelastic scatterings on nuclei via neutral currents, 
$\nu + (A,Z) \rightarrow \nu + (A,Z)^{*}$, 
the heating rate per unit volume, ${\cal Q}^{\alpha}$, is evaluated as
\begin{eqnarray}
 {\cal Q}^{\alpha}
  = \frac{\rho X_{A}}{m_{\rm B}}
  \frac{31.6\mev}{(r/10^{7}\text{cm})^{2}}
  &&\left[
     \frac{L_{\nu_{\text{e}}}}{10^{52}\text{ergs~s}^{-1}}
     \left(\frac{5\mev}{T_{\nu_{\text{e}}}}\right)
     \frac{A^{-1}\langle
     \sigma_{\nu_{\text{e}}}^{+}E_{\nu_{\text{e}}}
     +\sigma_{\nu_{\text{e}}}^{0}E_{\text{ex}}^{\text{A}}
     \rangle_{T_{\nu_{\text{e}}}}}
     {10^{-40}\text{cm}^2\mev} \right. \nonumber \\
 &&+\quad\frac{L_{\bar{\nu}_{\text{e}}}}{10^{52}\text{ergs~s}^{-1}}
  \left(\frac{5\mev}{T_{\bar{\nu}_{\text{e}}}}\right)
  \frac{A^{-1}\langle
  \sigma_{\bar{\nu}_{\text{e}}}^{-}E_{\nu_{\text{e}}}
  +\sigma_{\bar{\nu}_{\text{e}}}^{0}E_{\text{ex}}^{\text{A}}
  \rangle_{T_{\bar{\nu}_{\text{e}}}}}
  {10^{-40}\text{cm}^2\mev} \nonumber \\
 &&+\left.\quad
    \frac{L_{\nu_{x}}}{10^{52}\text{ergs~s}^{-1}}
    \left(\frac{10\mev}{T_{\nu_{x}}}\right)
    \frac{A^{-1}\langle
    \sigma_{\nu_{x}}^{0}E_{\text{ex}}^{\text{A}}
    +\sigma_{\bar{\nu}_{x}}^{0}E_{\text{ex}}^{\text{A}}
    \rangle_{T_{\nu_{x}}}}
    {10^{-40}\text{cm}^2\mev}
   \right],
 \label{eq:inelastic}
\end{eqnarray}
where $X_{\rm A}$ is the mass fraction of the nucleus
and $m_{\rm B}$ is the atomic mass unit~\citep{haxton88}.
The last term denotes the sum of the contributions from $\mu$ and $\tau$ neutrinos.
The cross section for each neutral-current is evaluated
by the following fitting formula,
\begin{equation}
 A^{-1}\langle
  \sigma_{\nu}^{0}E_{\text{ex}}^{\text{A}}
  +\sigma_{\bar{\nu}}^{0}E_{\text{ex}}^{\text{A}}
  \rangle_{T_{\nu}}
  = \alpha \left[ \frac{T_{\nu} - T_{0}}{10\mev} \right]^{\beta},
  \label{eq:fitting}
\end{equation}
where $\alpha$, $\beta$, and $T_{0}$ are given in Table~I of \citet{haxton88}, and are chosen to be
$\alpha=1.24\times 10^{-40}$~MeV~cm$^{2}$, $\beta=3.82$, and $T_{0}=2.54$~MeV~\citep{ohnishi07}.
In the first and second terms on the right hand side of Eq.~(\ref{eq:inelastic}),
the contributions from the charged current reactions,
$\sigma_{\nu}^{+}$ and $\sigma_{\nu}^{-}$,
are also taken into account according to Table~II of \citet{haxton88}, 
that is, $A^{-1}\langle \sigma_{\nu_{\text{e}}}^{+}E_{\nu_{\text{e}}} \rangle_{T_{\nu_{\text{e}}}} 
= 0.30 \times 10^{-42} \mev \cm^2$ for $T_{\nue} = 4\mev$
and $A^{-1}\langle \sigma_{\bar{\nu}_{\text{e}}}^{-}E_{\bar{\nu}_{\text{e}}} \rangle_{T_{\bar{\nu}_{\text{e}}}}
= 1.20 \times 10^{-42} \mev \cm^2$ for $T_{\nueb} = 5\mev$.

\subsection{The rate of change in specific energy}

The rate of change in energy in unit volume, $Q_{\rm E}$, is evalutated as
\begin{equation}
 Q_{\rm E} = (\Qae +\Qaa) +{\cal Q}^{\alpha}
  -(\Qee +\Qea) -3{\cal Q}^{\rm br}_{\nu\bar{\nu}}
-({\cal Q}^{\rm pa}_{\nue} +{\cal Q}^{\rm pa}_{\nueb} +4{\cal Q}^{\rm pa}_{\nux} )
-({\cal Q}^{\rm pl}_{\nue} +{\cal Q}^{\rm pl}_{\nueb} +4{\cal Q}^{\rm pl}_{\nux} ),
\end{equation}
where a factor of 3 in $3{\cal Q}^{\rm br}_{\nu\bar{\nu}}$ corresponds three types of a neutrino pair $\nu\bar{\nu}$.

\section{The rate of change in electron fraction, $Q_{\rm N}$}

The rate of change in $Y_e$ per unit volume, $Q_{\rm N}$, is evalutated as
\begin{equation}
 Q_{\rm N} = (\Rae -\Raa -\Ree +\Rea) \frac{m_{\rm B}}{\rho},
\end{equation}
where $\Rae$ and $\Raa$ are the absorption rates of $\nue$ and $\nuebar$, respectively,  
and $\Rea$ and $\Rea$ are the emission rates of $\nue$ and $\nueb$
through the capture of electrons on protons and that of positron on neutrons, respectively.
We ignore the variations of the electron fraction by the neutrino absorption on \nuc{He}{4}, 
since they are minor and give no qualitative difference to the dynamics~\citep{ohnishi07}.

The absorption rates of $\nue$ and $\nuebar$ per baryon are computed as
\begin{equation}
 \Rae = \sigma c \frac{L_{\nue} \, n_{\rm n}}{4\pi r^2 c f_{\nue}}
  \, \frac{ \langle\epsne^2\rangle + 2\Delta \langle\epsne\rangle + \Delta^2 }{\langle\epsne\rangle} 
  \, \Theta(\langle\epsne\rangle), 
\end{equation}
and
\begin{equation}
 \Raa = \sigma c \frac{L_{\nuebar} \, n_{\rm p}}{4\pi r^2 c f_{\nuebar}}
  \, \frac{ \langle\epsna^2\rangle^{\star} + 2\Delta \langle\epsna\rangle^{\star} +
  \Delta^2 \langle\epsna^0\rangle^{\star}}{\langle\epsna\rangle},
\end{equation}
respectively.

The emission rate of $\nue$ per baryon through the electron capture on protons is given by
\begin{equation}
 \Ree = \frac{1}{2} \sigma c \, n_{\rm p} \, n_{{\rm e}^-} \,
  [ \langle\epsem^2\rangle^{\star} + 2\Delta \langle\epsem\rangle^{\star} 
  + \Delta^2 \langle\epsem^0\rangle^{\star} ]
\end{equation}
and that of $\nueb$ via the positron capture on neutrons by
\begin{equation}
 \Rea = \frac{1}{2} \sigma c \, n_{\rm n} \, n_{{\rm e}^+} \,
  [ \langle\epsep^2\rangle + 2\Delta \langle\epsep\rangle +\Delta^2 ].
\end{equation}

\end{appendix}

\end{document}